\documentclass[prl,aps,superscriptaddress,footinbib,twocolumn,floatfix]{revtex4-2}
\usepackage[latin9]{inputenc}
\usepackage{color}
\usepackage{amsmath}
\usepackage{multirow}
\usepackage{amssymb}
\usepackage{stmaryrd}
\usepackage{graphicx}
\usepackage[unicode=true,bookmarks=true,bookmarksnumbered=false,bookmarksopen=false,breaklinks=false,pdfborder={0 0 1},backref=false,colorlinks=true]
 {hyperref}
\hypersetup{
 linkcolor=magenta, urlcolor=blue, citecolor=blue, pdfstartview={FitH}, hyperfootnotes=false, unicode=true}

\makeatletter
\newcommand{\mcol}{\multicolumn{2}{c}}
\newcommand{\cfill}{\cellcolor[HTML]{44cef6}}
\newcommand{\cfillo}{\cellcolor[HTML]{f05654}}
\usepackage[table,xcdraw]{xcolor}
\usepackage{framed}
\usepackage{fbox} 
\usepackage{amsfonts}\usepackage{tabularx}\usepackage{dcolumn}\usepackage{bm}\usepackage{graphicx}\usepackage{epstopdf}
\usepackage{times}
\hypersetup{urlcolor=blue}

\def\ket#1{\left|#1\right\rangle}
\def\bra#1{\left\langle#1\right|}

\usepackage{tikz}
\usetikzlibrary{quantikz}
\usepackage{algorithm}
\usepackage{algorithmic}

\makeatother

\begin{document}
\title{Experimental quantum adversarial learning with programmable superconducting qubits}

\affiliation{Department of Physics, ZJU-Hangzhou Global Scientific and Technological Innovation Center, Interdisciplinary Center for Quantum Information, and Zhejiang Province Key Laboratory of Quantum Technology and Device, Zhejiang University, Hangzhou 310027, China\\
$^2$ Center for Quantum Information, IIIS, Tsinghua University, Beijing 100084, China\\
$^3$ Alibaba-Zhejiang University Joint Research Institute of Frontier
Technologies, Hangzhou 310027, China\\
$^{4}$ Skolkovo Institute of Science and Technology, Moscow 121205, Russia\\
$^{5}$ Shanghai Qi Zhi Institute, 41th Floor, AI Tower, No. 701 Yunjin Road, Xuhui District, Shanghai 200232, China\\
$^{6}$ State Key Laboratory of Modern Optical Instrumentation, Zhejiang University, Hangzhou 310027, China\\
}

\author{Wenhui Ren$^{1, *}$}
\author{Weikang Li$^{2, *}$}
\author{Shibo Xu$^{1, *}$}

\author{Ke Wang$^{1}$}
\author{Wenjie Jiang$^{2}$}
\author{Feitong Jin$^{1}$}
\author{Xuhao Zhu$^{1}$}

\author{Jiachen Chen$^{1}$}
\author{Zixuan Song$^{1}$}
\author{Pengfei Zhang$^{1}$}

\author{Hang Dong$^{1}$}
\author{Xu Zhang$^{1}$}
\author{Jinfeng Deng$^{1}$}

\author{Yu Gao$^{1}$}
\author{Chuanyu Zhang$^{1}$}
\author{Yaozu Wu$^{1}$}

\author{Bing Zhang$^{3}$}

\author{Qiujiang Guo$^{1, 3}$}
\author{Hekang Li$^{1, 3}$}
\author{Zhen Wang$^{1, 3}$}

\author{Jacob Biamonte$^{4}$}

\author{Chao Song$^{1, 3, \dagger}$}
\author{Dong-Ling Deng$^{2, 5, \S}$}
\author{H. Wang$^{1, 3, 6\ddagger}$}

\begin{abstract}
\textbf{
Quantum computing promises to enhance machine learning and artificial intelligence \cite{Biamonte2017Quantum,Dunjko2018Machine,Sarma2019Machine}. Different quantum algorithms have been proposed to improve a wide spectrum of machine learning tasks \cite{Gao2018Quantum,Liu2021Rigorous,Havlicek2019Supervised,Schuld2019Quantum,Saggio2021Experimental,Dunjko2016QuantumEnhanced,Peters2021Machine,Gong2022Quantum,Herrmann2021Realizing}. Yet, recent theoretical works show that, similar to traditional classifiers based on deep classical neural networks, quantum classifiers would suffer from the vulnerability problem: adding tiny carefully-crafted perturbations to the legitimate original data samples would facilitate incorrect predictions at a notably high confidence level \cite{Lu2020Quantum,Liu2020Vulnerability,Gong2021Universal,Guan2021Robustness,Liao2021Robust}. This will pose serious problems for future quantum machine learning applications in safety and security-critical scenarios \cite{Biggio2018Wild,Vorobeychik2018Adversarial,Finlayson2019Adversarial}. Here, we report the first experimental demonstration of quantum adversarial learning with programmable superconducting qubits. We train quantum classifiers, which are built upon variational quantum circuits consisting of ten transmon qubits featuring average lifetimes of 150~$\mu$s, and average fidelities of simultaneous single- and two-qubit gates above 99.94\% and 99.4\% respectively, with both real-life images (e.g.,  medical magnetic resonance imaging scans) and quantum data. We demonstrate that these well-trained classifiers (with testing accuracy up to 99\%) can be practically deceived by small adversarial perturbations, whereas an adversarial training process would significantly enhance their robustness to such perturbations. Our results  reveal experimentally a crucial vulnerability aspect of quantum learning systems under adversarial scenarios and demonstrate an effective defense strategy against adversarial attacks, which  provide a valuable guide for quantum artificial intelligence applications with both near-term and future quantum devices. 
 }
\end{abstract}

\maketitle
In recent years, artificial intelligence (AI) \cite{Silver2016Mastering,Jumper2021Highly,Davies2021Advancing} and quantum computing \cite{Arute2019Quantum,Wu2021Strong,Gong2021Quantum} have made dramatic progress. Their intersection gives rise to a research frontier called, quantum machine learning or generally, quantum AI \cite{Biamonte2017Quantum,Dunjko2018Machine,Sarma2019Machine}. A number of quantum algorithms have been proposed to enhance various AI tasks \cite{Gao2018Quantum,Liu2021Rigorous,Havlicek2019Supervised,Schuld2019Quantum,Saggio2021Experimental,Dunjko2016QuantumEnhanced,Peters2021Machine,Gong2022Quantum,Herrmann2021Realizing}.  With the rapid establishment of quantum enhanced AI, a pressing, fundamental question emerges naturally: are quantum AI technologies trustworthy under adversarial attacks? 

Classical neural networks are vulnerable to adversarial perturbations. For instance, a stop sign with small graffiti might be misclassified as a yield sign \cite{Eykholt2018Robust}, whereas adding a tiny amount of carefully-crafted noise---which is even imperceptible 
to the human eye---into an image of a benign skin lesion would fool the classifier to predict it as malignant \cite{Finlayson2019Adversarial}. This surprising vulnerability of classical neural networks has far-reaching consequences in safety and security-critical scenarios (e.g., autonomous driving, biometric authentication, and medical diagnostics). More recently, the vulnerability of quantum classifiers has been studied, establishing the foundations of quantum adversarial machine learning \cite{Lu2020Quantum,Liu2020Vulnerability,Gong2021Universal,Guan2021Robustness,Liao2021Robust}. It has been shown theoretically that quantum classifiers are likewise highly vulnerable to adversarial examples, independent of the learning algorithms and regardless of whether the input data is classical or quantum \cite{Lu2020Quantum}. In addition, different countermeasures, such as adversarial training \cite{Kurakin2017Adversarial}, have also been proposed to enhance the robustness of quantum classifiers against adversarial perturbations. However, demonstrating adversarial examples for quantum classifiers experimentally and showing the effectiveness of the proposed countermeasures in practice are challenging and have not previously been reported. To accomplish this, one faces at least two difficulties: (i) determining an experimentally feasible encoding of high-dimensional classical data, and (ii) building quantum classifiers with a large enough state-space so as to identify realistic images. 

\begin{figure*}[tbp]
\center
\includegraphics[width=1.0\linewidth]{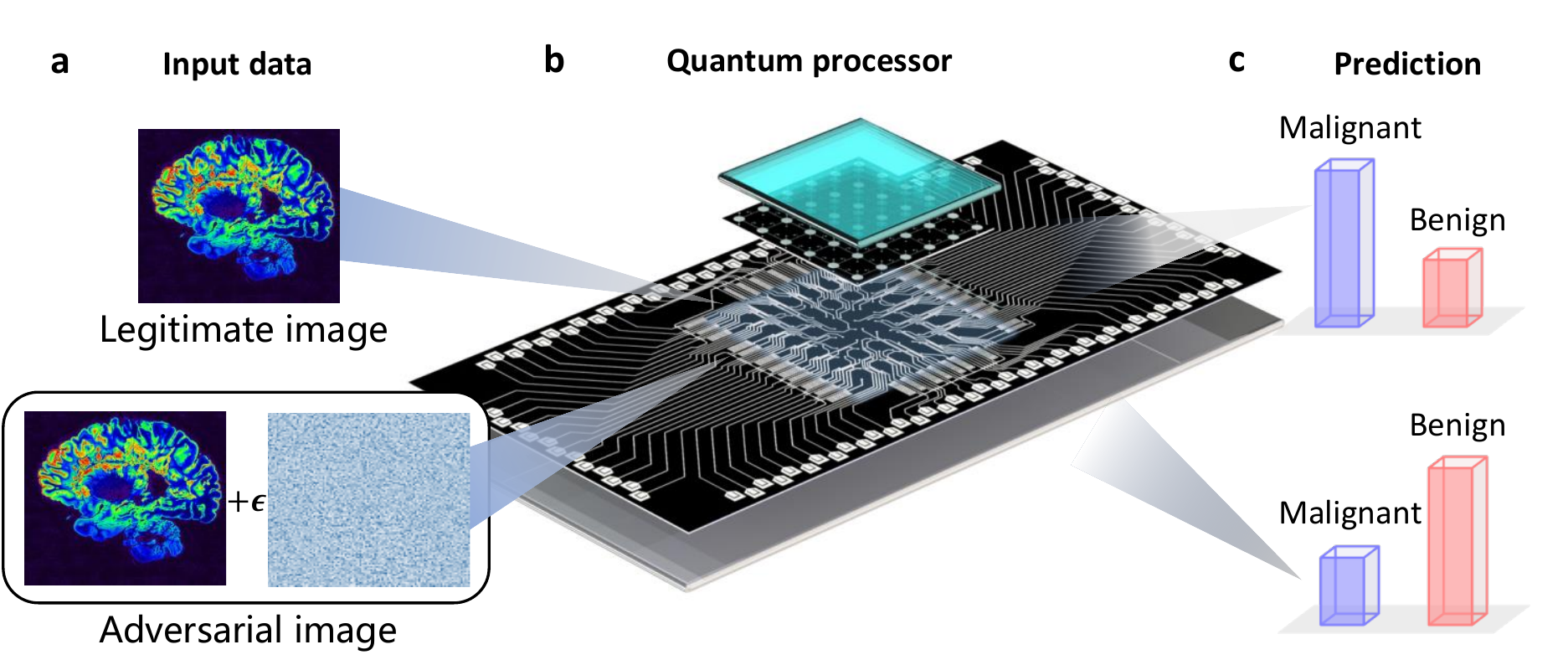}
\caption{\textbf{Schematic of experimental quantum adversarial learning.}
\textbf{a}, A legitimate MRI (magnetic resonance imaging) scan of a fixed cerebral hemisphere for sclerosis diagnosis \cite{MRIscan} and its corresponding adversarial example, which is obtained by adding a tiny amount of carefully-crafted perturbations to the original image.
\textbf{b}, Exhibition of a programmable quantum processor with $36$ superconducting transmon qubits arranged on a $6\times6$ square lattice. The qubit layer and control-line layer as highlighted are patterned on the sapphire (top) and silicon (bottom) substrates respectively, which are assembled together during the flip-chip bonding process. The quantum classifiers are build upon large-scale variational quantum circuits implemented with this processor. \textbf{c}, Predictions for the legitimate and adversarial samples. The quantum classifier will correctly identify the legitimate MRI scan as ``Malignant", whereas incorrectly classify the corresponding adversarial example, which differs by only an imperceptible amount of perturbation, into the ``Benign" class with a high confidence.}
\label{fig1}
\end{figure*}

Here, we overcome these difficulties and report the first experimental demonstration of quantum adversarial learning with an array of ten programmable superconducting transmon qubits. Through optimizing device fabrication and controlling process, we push the average lifetime of these qubits to 150~$\mu$s and the average simultaneous single- and two-qubit gate fidelities greater than 99.94\% and 99.4\%, respectively. This enables us to successfully implement large-scale quantum classifiers with different structures up to a circuit depth of $60$ and the number of trainable variational parameters exceeding $250$. We train these classifiers with both large-size real-life images (e.g., medical magnetic resonance imaging scans \cite{MedicalMNIST,Clark2013Cancer,Halabi2019RSNA,Wang2017ChestXray8}) and high-dimensional quantum data (e.g., thermal and localized quantum many-body states), through quantum gradients obtained directly by measuring some observables. After training, these classifiers can achieve the state-of-the-art performance on these datasets, with a testing accuracy up to 99\%. We generate adversarial examples through a classical optimizing procedure and show unambiguously that they can deceive the trained quantum classifiers with a high confidence level. To mitigate such vulnerability, we further demonstrate that, through adversarial training, the quantum classifiers will be immune to adversarial perturbations generated by the same attacking strategy.

\vspace{.5cm}
\noindent\textbf{\large{}Framework and experimental setup}{\large\par}

\noindent We first introduce the general framework for quantum adversarial machine learning. We consider classification tasks in the setting of supervised learning \cite{Lu2020Quantum}, where we train quantum classifiers with pre-labeled data samples, through minimizing the following loss function iteratively
\begin{eqnarray}\label{CE}
\mathcal{L}\left(h\left(\mathbf{x} ; \boldsymbol{\theta}\right), \mathbf{a}\right)=-\sum_{k} a_{k} \log g_{k}.
\end{eqnarray}
Here, $\mathbf{x}$ denotes a training sample, $h(\mathbf{x};\boldsymbol{\theta})$ represents the hypothesis function determined by the quantum classifier with variational parameters denoted collectively as $\boldsymbol{\theta}$, $\mathbf{a}$ is the one-hot encoding of the labels, and $g_k$ denotes the probability for the $k$-th category obtained from measuring the quantum classifier (see Methods). After the training process, the quantum classifier will typically be able to assign labels to data samples outside the training set with high accuracy. To obtain adversarial examples, we focus on the scenario of untargeted white-box attacks, where we assume the attacker has full information about the quantum classifier and no particular class is aimed \cite{Lu2020Quantum}. Unlike the training process, where we vary the variational parameters to minimize the loss, for generating adversarial examples we fix $\boldsymbol{\theta}$ at its optimal value $\boldsymbol{\theta}^*$ obtained at the last step of the training, and optimize over the input space within a small region to maximize the loss function instead (see Methods and the Supplementary Sec.~IB). We input the generated adversarial examples into the quantum classifier to test its performance. A schematic illustration of the main idea for quantum adversarial learning is shown in Fig.~\ref{fig1}.

\begin{figure*}[tbp]
\center
\includegraphics[width=1\linewidth]{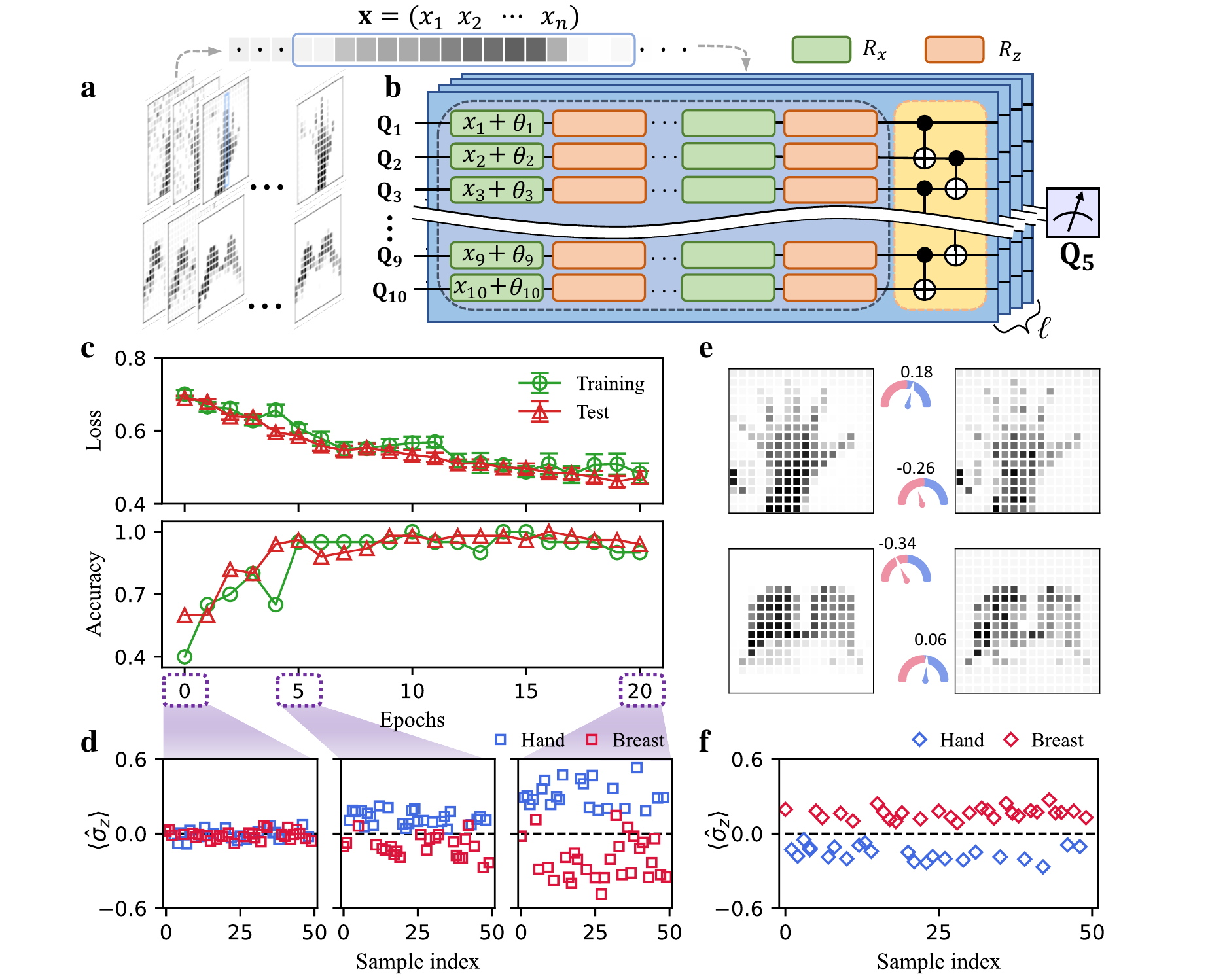}
\caption{\textbf{The framework of a quantum neural network for learning medical data and the experimental demonstration of its vulnerability to adversarial perturbations.} \textbf{a}, Encoding of the medical hand-breast MRI data. We compress each MRI image to $16\times16$ pixels, which is represented by a $256$-dimensional vector $\mathbf{x}$ encoded into the quantum neural network classifier. \textbf{b}, Experimental quantum circuit to realize the interleaved block-encoding quantum classifier. The circuit is composed of $l$ blocks, of which each consists of multiple layers of single-qubit rotational gates (dash blue box) followed by two layers of CNOT gates (dash yellow box). The rotation angles are obtained by summing up $\mathbf{x}$ and variational parameters $\boldsymbol{\theta}$. \textbf{c}, Loss function (up) and accuracy (down) for the training and testing dataset at each epoch during the training process of the quantum classifier. \textbf{d}, Experimentally measured $\langle\hat{\sigma}_z\rangle$ of Q$_5$ for the test (square) data at epoch 0, 5 and 20 respectively. Data points for samples labeled ``hand'' and ``breast'' are colored in blue and red, respectively. \textbf{e}, Legitimate and adversarial samples with measured output $\langle\hat{\sigma}_z\rangle$ for Q$_5$ of the trained quantum classifier. \textbf{f},  Experimentally measured $\langle\hat{\sigma}_z\rangle$ for adversarial examples when input into the trained quantum classifier (at the epoch $20$ of the training process).}
\label{fig2}
\end{figure*}

Our experiment is implemented on a flip-chip superconducting quantum processor, which possesses 36 transmon qubits arranged in a two-dimensional array featuring tunable nearest neighbor couplings (Fig.~\ref{fig1}\textbf{b}). To achieve high coherence we deposited tantalum films, using a high-vacuum sputtering system (Yunmao QBT-P), which were patterned for qubit structures. For the purpose of demonstrating quantum adversarial learning, we choose a one-dimensional array of ten qubits, whose energy relaxation times $T_1$ range from 131 to 173~$\mu$s at the frequencies where the qubits are initialized and operated.
Single-qubit XY rotations are realized using $30$ ns-long microwave pulses which are generated by multi-channel arbitrary waveform generators (MOSTFIT MF-AWG-08), and the controlled-NOT (CNOT) gate is based on controlled-$\pi$ phase (CZ) gate plus single-qubit rotations.
The CZ gate, which has a length of $60$ ns, is realized by carefully tunning the frequencies and coupling strength of qubits to steer a closed-cycle diabatic transition of $|11\rangle\leftrightarrow|20\rangle$ (or $|02\rangle $)~ \cite{PhysRevX.11.021058, PhysRevLett.125.120504}. Since single-qubit (two-qubit) gates are simultaneously implemented on multiple qubits (qubit pairs) in our experimental sequences, we carry out simultaneous cross entropy benchmarkings to characterize gate performances, yielding average \emph{Pauli} error around $0.08\%$ ($0.72\%$). See Supplementary Sec.~III for details on device and gate performances.

\begin{figure*}[tbp]
\center
\includegraphics[width=0.9\linewidth]{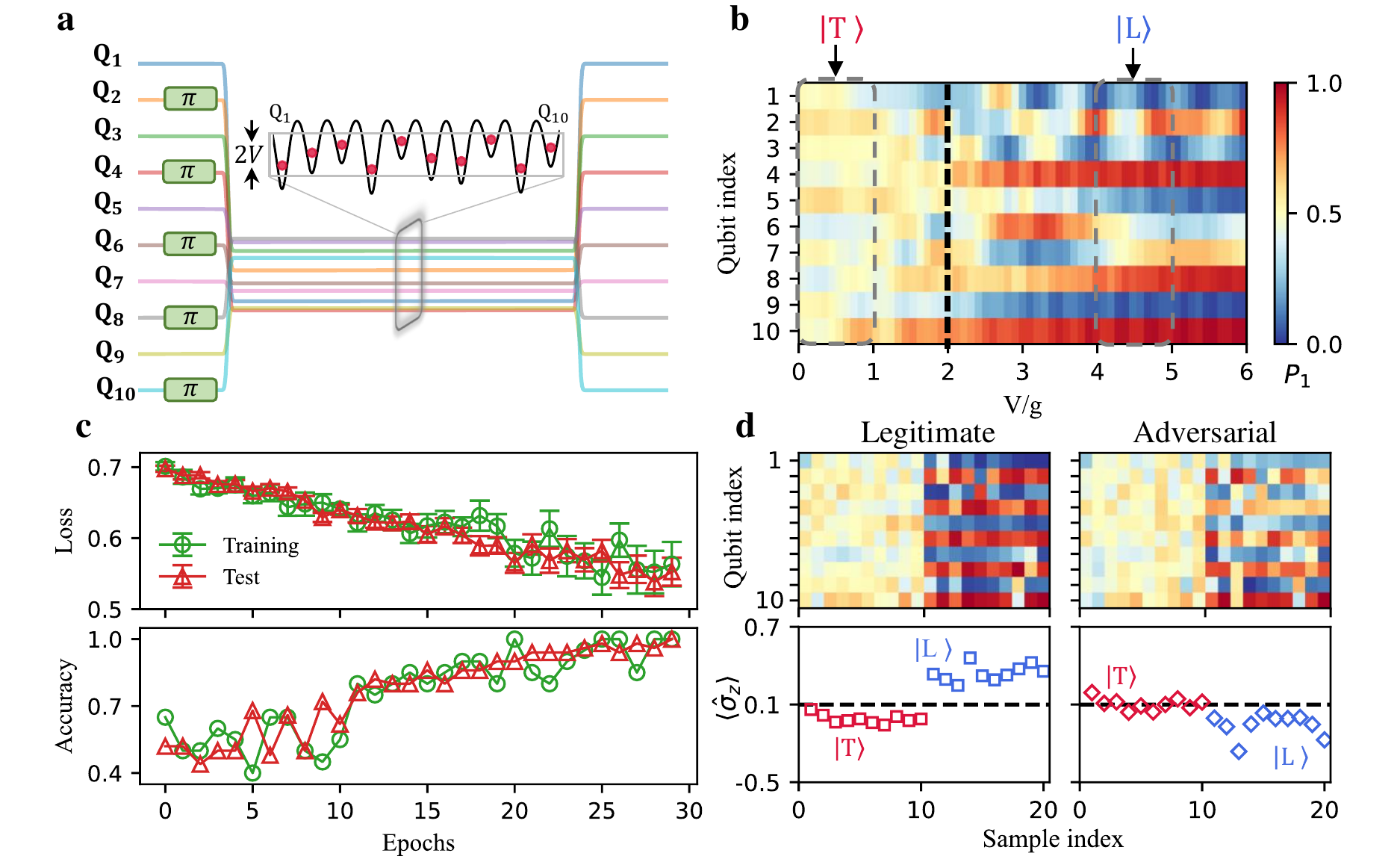}
\caption{\textbf{Experimental results for learning quantum data.} \textbf{a}, Pulse sequences for generating quantum data. After preparing the system into the N\'{e}el state, we tune the frequency of each qubit to engineer the incommensurate potential of the Aubry-Andr\'{e} model (inset) and wait for 400 ns until the system evolves into the desired state. \textbf{b}, The excited state probability $P_1$ of each qubit with $\phi$ fixed to 0 and different $V/g$. 
The system hosts a transition from thermal to localized phases at the critical point $V/g=2$ (dash line). 
The two categories of quantum data, labeled as $|\text{T}\rangle$ (thermal) and $|\text{L}\rangle$ (localized), are sampled from $V/g\in [0,1]$ and $[4,5]$ (gray boxes) respectively with random $\phi$. \textbf{c}, Loss function (up) and accuracy (down) for the test and training set at each epoch. \textbf{d}, Vulnerability of the quantum classifier in learning quantum states. We select ten legitimate $|\text{T}\rangle$ and $|\text{L}\rangle$ states from the training set, whose local magnetization distribution and classification outputs are shown in the top and lower left panels. After applying adversarial perturbations on the legitimate states, half of the $|\text{T}\rangle$ and all the $|\text{L}\rangle$ states are classified incorrectly by the trained classifier (lower right), even though the essential features of local magnetization distribution (top right) are still clearly distinct for thermal and localized regions.}
\label{fig3}
\end{figure*}

\vspace{.5cm}
\noindent \textbf{\large{}Quantum adversarial learning medical data}{\large\par}

\noindent Machine learning has cemented its role in modern medical related technologies, ranging from the development of healthcare systems \cite{Friedman2010Achieving}, and the dermatologist-level classification of skin cancer \cite{Esteva2017Dermatologist}, to the prediction of the progression from pre-dabetes to type II diabetes using routinely-collected  health record data \cite{Anderson2016Reverse}. This is such a safety and security-critical area, where incorrect predictions of the learning system may cost billions of dollars for healthcare insurance companies or even lead to possible medical disasters \cite{Finlayson2019Adversarial}. Quantum machine learning holds vast potential in medical applications. Yet, similar to the classical medical learning, its possible vulnerabilities likewise demand careful study.

To investigate the vulnerability of quantum learning systems in medical diagnoistics, we consider a binary classification task for identifying MRI images. We exploit an interleaved block-encoding theme \cite{Caro2021Encoding,Haug2021Largescale,Perez-Salinas2020Data}, rather than the conventional amplitude encoding, to encode the input classical data (Methods and Supplementary Sec.~IA). This enables us to circumvent the notorious difficulty of preparing a highly-entangled multiqubit quantum state and is crucial for the success of classifying large-size images (16 by 16 pixels, Fig.~\ref{fig2}\textbf{a}) by a large-scale quantum classifier (up to 260 trainable parameters) with the state-of-the-art (but still rather limited) gate fidelities. The interleaved block-encoding and the structure of our quantum classifier are illustrated in Fig.~\ref{fig2}\textbf{b}. We train our quantum classifier with MRI images labeled by ``Hand" and ``Breast", through quantum gradients obtained directly by measuring some observables in our experiment (Supplementary Sec.~IA).  Our experimental result for the training process is plotted in Fig.~\ref{fig2}\textbf{c}, from which it is clear that the accuracy for both the training and test datasets increases rapidly at the beginning of the training process and then saturate at a high value ($0.92$ and $0.97$ for the training and test datasets, respectively).  In Fig.~\ref{fig2}\textbf{d}, we plot the measured $\langle\hat{\sigma}_z\rangle$ value, which determines the assigned labels (``Hand" and ``Breast" for $\langle\hat{\sigma}_z\rangle\geq 0$ and $\langle\hat{\sigma}_z\rangle< 0$, respectively),  for samples from the test dataset at different iteration steps. We find that at the beginning (left subfigure), $\langle\hat{\sigma}_z\rangle$ concentrates near zero, which agrees with the fact that the variational parameters for the quantum classifier are randomly initialized. After five training epochs (middle subfigure), $\langle\hat{\sigma}_z\rangle$ become clearly bifurcated ($\langle\hat{\sigma}_z\rangle>0$ for all the ``Hand" images and $\langle\hat{\sigma}_z\rangle<0$ for most of the ``Breast" images), resulting in a test accuracy of about $0.96$. After $20$ epochs (right subfigure), the bifurcation of $\langle\hat{\sigma}_z\rangle$ is larger, which is consistent with the decrease of the loss function as shown in Fig.~\ref{fig2}\textbf{c} (up panel).

After the training process, we fix the variational parameters of the quantum classifier and solve the following optimization problem to obtain adversarial perturbations (which are then added to the corresponding legitimate MRI images to generate adversarial examples, see Supplementary Sec.~IB)
\begin{eqnarray}
    \boldsymbol{\delta} \equiv \underset{\boldsymbol{\delta^{\prime}} \in \Delta}{\operatorname{argmax}}\; \mathcal{L}\left(h\left(\mathbf{x}+\boldsymbol{\delta^{\prime}} ; \boldsymbol{\theta}^{*}\right), \mathbf{a}\right),
\end{eqnarray}
where $\Delta$ denotes a small region introduced to ensure that the adversarial perturbations are small and will not alter the input data essentially. In Fig.~\ref{fig2}\textbf{e}, we plot the original legitimate MRI images (left column) and their corresponding adversarial ones (right column), together with their measured $\langle\hat{\sigma}_z\rangle$ values. From this figure, we see that the adversarial images differ from the legitimate ones only by a tiny amount of perturbations (almost imperceptible to human eyes), yet the well-trained quantum classifier will assign incorrect labels to them, as indicated by the corresponding measured $\langle\hat{\sigma}_z\rangle$ values.  In addition, Fig.~\ref{fig2}\textbf{f} shows  $\langle\hat{\sigma}_z\rangle$ for all adversarial examples corresponding to the original MRI images in the test set. We find that the quantum classifier misclassifies all of them. This 
unambiguously manifests the vulnerability aspect of quantum classifiers in learning medical images.

\vspace{.5cm}
\noindent\textbf{\large{}Adversarial examples for quantum data}{\large\par}

\noindent Unlike classical classifiers that can only take classical data as input, quantum classifiers can also naturally handle quantum states as input and gain potential exponential advantages. We now show the vulnerability of quantum classifiers in classifying quantum states. For concreteness, we consider a binary classification of quantum states generated by evolving the N\'{e}el state for a period of time with the following Aubry-Andr\'{e} Hamiltonian \cite{Aubry1980Analyticity}:
\begin{eqnarray}\label{AAH}
    H/\hbar = -\frac{g}{2}\sum_{k}(\hat{\sigma}_{k}^{x}\hat{\sigma}_{k+1}^{x}+\hat{\sigma}_{k}^{y}\hat{\sigma}_{k+1}^{y}) - \sum_{k}\frac{V_k}{2}\hat{\sigma}_{k}^{z},
\end{eqnarray}
where $g$ is the coupling strength, $\hat{\sigma}_{k}^{l}$ ($l=x,y,z$) 
is the Pauli operator for the $k$-th qubit, and $V_k=V\cos(2\pi\alpha k+\phi)$ is the incommensurate potential with $V$ being the disorder magnitude, 
$\alpha=(\sqrt{5}-1)/2$ being an irrational number and $\phi$ being a random phase evenly distributed on $[0,2\pi)$. 
This Hamiltonian features a quantum phase transition at $V/g=2$, between a localized phase for $V/g>2$ and a delocalized (thermal) phase for $V/g<2$ \cite{Aubry1980Analyticity}. In our experiment, we initialize the system to the N\'{e}el state and then evolve it under $H$ for about $400$ ns, with the pulse sequence sketched in Fig.~\ref{fig3}\textbf{a}.
We fix $g/2\pi\approx 5$~MHz and scan $V/2\pi$ from $0$~MHz to $30$~MHz. In Fig.~\ref{fig3}\textbf{b}, we  plot the measured probability $P_1$ of being on state $|1\rangle$ for each qubit (equivalent to the  local magnetization $\langle\hat{\sigma}_z\rangle$ by noting $P_1\equiv \frac{1}{2}-\frac{1}{2}\langle\hat{\sigma}_z\rangle$)  for varying $V$, from which the localized and thermal features of the evolved states are clearly manifested. 

We randomly choose some of the evolved quantum states deep in the localized and thermal regions (dashed grey boxes in Fig.~\ref{fig3}\textbf{b}) to form a quantum dataset. We implement a quantum classifier, which consists of five layers with each containing three single-qubit rotations and two controlled-NOT gates, to classify the chosen states in a supervised fashion (Supplementary Sec.~II). We randomly initialize the $150$ variational parameters and train the quantum classifier with experimentally obtained quantum gradients. Fig.~\ref{fig3}\textbf{c} plot the accuracy and loss as a function of epochs obtained in our experiment during the training process. We find that the implemented quantum classifier has an excellent performance in this task and after about $30$ iteration steps it achieves near perfect  accuracy on both the training and test datasets.   

Similar to the case of learning medical images, the quantum classifier is vulnerable to adversarial perturbations in leaning quantum states as well.  To demonstrate this in our experiment, we generate adversarial perturbations for state samples in the test set by solving an optimization problem with quantum gradients measured in experiment (Methods and Supplementary Sec.~IIB). We add the obtained perturbations to their corresponding legitimate states through adding a near-identity unitary before input the states into the quantum classifier. In the first row of Fig.~\ref{fig3}\textbf{d}, we randomly choose 20 states from the training set and plot their measured  $P_1$ values of each qubits, for both the legitimate (left) and adversarial (right) samples. From this figure, the adversarial examples differs slightly from the legitimate ones (especially for these in the thermal region, the difference is indiscernibly small) and maintain the essential features (i.e., vanishing and persistent local magnetization) for thermal and localized states, respectively. However, they would successfully deceive the quantum classifier with very large probability, as indicated in the second row of Fig.~\ref{fig3}\textbf{d}. From the left subfigure, it is clear that the trained classifier can correctly identify all the legitimate states. Whereas, it will misclassify all (half) of the adversarial examples in the localized (thermal) region, as shown in the right subfigure. This demonstrates lucidly the vulnerability of quantum classifiers to adversarial perturbations in categorizing quantum states.

\begin{figure}[tbp]
\center
\includegraphics[width=1.0\linewidth]{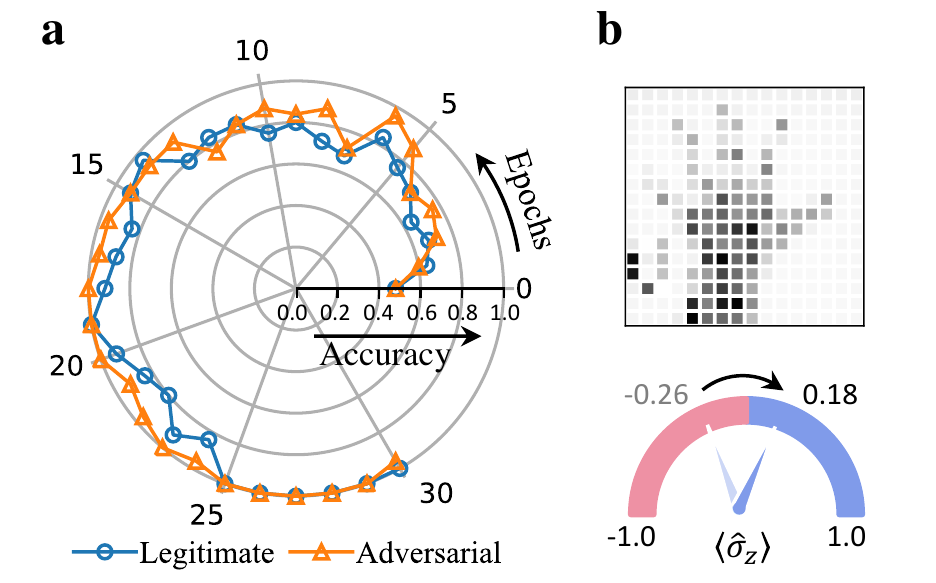}
\caption{\textbf{Experimental results for quantum adversarial training with MRI images.} \textbf{a}, Accuracy for the legitimate and adversarial test data at each epoch during the adversarial training process. \textbf{b}, An image of an adversarial sample and the corresponding experimental outputs before and after adversarial training of the quantum classifier. Before adversarial training, the classifier will misclassify this sample as ``Breast" (as indicated by the output $\langle\hat{\sigma}_z\rangle=-0.26$),  whereas after adversarial training, it will restore its validity and identify the sample correctly as ``Hand" again (as indicated by $\langle\hat{\sigma}_z\rangle=0.18$).
}
\label{fig4}
\end{figure}

\vspace{.5cm}
\noindent \textbf{\large{}Adversarial training of quantum classifiers}{\large\par}

\noindent In the above discussion, we have shown with concrete examples that quantum learning systems are rather fragile to adversarial attacks. This may lead to severe problems for their applications, especially for these in safety and security-sensitive scenarios, ranging from autonomous driving \cite{Bojarski2016End} and medical diagnoistics \cite{Finlayson2019Adversarial} to quantum finance \cite{Orus2019Quantum} and biometric authentication \cite{Bhattacharyya2009Biometric}. In theory, a variety of defense strategies have been proposed to enhance the robustness of quantum learning systems against adversarial perturbations, including adversarial training \cite{Lu2020Quantum} and exploiting quantum noises \cite{Du2021Quantum}. 

Here, we focus on adversarial training and carry out an experiment to demonstrate its effectiveness in practice. We first numerically generate adversarial examples for each legitimate sample and then inject them into the training set. We retrain the quantum classifier with both the legitimate and adversarial samples (Methods). In Fig.~\ref{fig4}\textbf{a}, we plot the accuracy of the classifier for classifying MRI images on both the legitimate and adversarial sets, as a function of epochs during the adversarial training process. We find that it increases for both datasets and approaches unity after about $25$ epochs, indicating that the adversarially retrained quantum classifier becomes immune to adversarial perturbations. To be more concrete, in Fig.~\ref{fig4}\textbf{b} we plot a randomly chosen adversarial example (up panel). This image will be misclassified by the original quantum classifier into the category of ``Breast" (with $\langle\hat{\sigma}_z\rangle=-0.26$), yet after adversarial training it will be identified correctly as ``Hand" (with a refreshed $\langle\hat{\sigma}_z\rangle$ value of $0.18$). This shows explicitly that adversarial training can indeed significantly enhance the robustness of quantum classifiers against adversarial perturbations.

\vspace{.5cm}
\noindent \textbf{\large{}Conclusions and outlook}{\large\par}
\noindent Theoretically, the existence of adversarial examples has an origin in the fundamental concentration of measure phenomenon \cite{ledoux2001concentration} and is hence an inevitable feature for quantum machine learning with high-dimensional data \cite{Lu2020Quantum,Liu2020Vulnerability,Gong2021Universal}, independent of the learning models, the training algorithms, and whether the input data is classical or quantum. In this work, our discussion is mainly focused on supervised learning based on quantum circuit classifiers. The experimental demonstration of quantum adversarial examples for unsupervised learning and other types of quantum classifiers \cite{Li2022Recent} seems more technically sophisticated and still remain unattainable. In addition, other defense strategies such as defensive distillation \cite{Papernot2016Distillation} and defense-GAN (generative adversarial network) \cite{Samangouei2018Defense} have also been introduced in the classical adversarial machine learning literature. It would be interesting and important to extend these strategies to the quantum domain, both in theory and experiment. In particular, we note that a quantum version of GAN (qGAN) has already been demonstrated experimentally \cite{Hu2019Quantum,Huang2021Quantum}.  Yet, how to construct a defense-qGAN that would substantially enhance the robustness of quantum learning systems to adversarial perturbations and how to implement it in experiment remain still unclear and worth further investigation.

Undoubtedly, the promise of quantum AI is huge. Yet, how to build a trustworthy quantum AI system and deliver this promise to practical applications remains largely unclear and demands long-term research. Our results make a crucial experimental attempt towards trustworthy quantum AI by not only revealing the vulnerability of quantum learning systems in adversarial scenarios, but also demonstrating the effectiveness of a defense strategy against adversarial attacks in practice. As the fledgling field of quantum AI grows, our results will prove useful in practical applications that are safety and security critical.

\vspace{.5cm}
\noindent\textbf{\large{}Methods}{\large\par}

\noindent\textbf{Quantum classifiers with classical data}

\noindent Here, we introduce the detailed settings of the quantum classifier for the classical dataset.
The quantum classifier is composed of several blocks, while each block contains several layers of single qubit gates and ends with two layers of CNOT gates that entangle all the qubits.
For each block, as shown in Fig.~\ref{fig2}\textbf{b}., the single qubit gates can be utilized to encode both trainable parameters and the input data.
To encode the image information from the medical MRI dataset \cite{MedicalMNIST,Clark2013Cancer,Halabi2019RSNA} into the quantum classifier, we first compress the images down to $16$ by $16$ pixels,
which are then normalized and mapped into the rotation angles of the single qubit gates in the quantum classifier by a factor of two.
For concreteness, since we are using a ten-qubit quantum classifier, 
we use $26$ layers of single-qubit variational gates to encode the $256$-dimensional data by adding four ``0'' at the end of the data vectors.
For each rotation angle that encodes the input samples, we attach one trainable parameter that can be optimized with gradient descent methods.

For the hyperparameter setting of the experimental demonstrations, we select the ``Hand'' and ``Breast'' MRI images from the medical dataset.
The size of the training set and the test set are $500$ and $100$ , respectively.
To measure the distance between the current output and the target label, 
we choose cross entropy as the loss function (Eq.~\ref{CE}), 
and the learning rate is set to be $0.05$.

The quantum classifier is initialized with randomly generated trainable parameters. 
During the training process, we divide the $260$ trainable parameters into ten groups.
For each epoch, we update the parameters in these groups sequentially.
To train the parameters in each group, we randomly select $20$ ($50$) samples from the training (test) set,
where the $20$ samples from the training set are utilized to calculate the gradients and optimize the parameters in the classifier, and the $50$
test samples are utilized to approximately calculate the test accuracy.
The loss function and accuracy of both training and test data measured at each epoch are plotted in Fig.~\ref{fig2}\textbf{c}. 
As the loss function decreases slowly during the learning process, 
the accuracy increases at a relatively faster speed and approaches to  saturated values after about five epochs. 
Further decrease of the loss function helps to enhance the separation between the two categories, as witnessed by the instances in Fig.~\ref{fig2}\textbf{d}. 
After $20$ epochs,
the trained quantum classifier is able to classify the total training (test) set with accuracy $0.92$ ($0.97$).
We note that, to minimize the circuit depth in order to reduce the experimental noise, 
we recompile the quantum circuit before the actual execution by replacing the single qubit gates with two gates,
i.e., $R_{\phi}(\alpha)$ and $R_{z}(\theta)$ (Supplementary Sec.~IIIB). 
Moreover, dynamical decoupling pulses are applied on the qubits during their idling times in the quantum circuits.

We mention that, in addition to the learning task for the medical data in the main text, we have demonstrated the quantum adversarial learning of MNIST handwritten digit dataset \cite{LeCun1998Mnist} as well to exam the feasibility of our protocol.
For this task,
The basic quantum circuit settings are the same as that for the medical dataset,
and the images of digits ``$0$'' and ``$1$'' are selected to form the training and test set.
For experimental convenience, we only choose $50$ of these parameters to be trained, which lie at the $3$rd, $6$th, $11$th, $17$th, $23$rd single-qubit layers of the quantum classifier.
The experimental results for learning MNIST handwritten digit dataset are shown in Supplementary Sec.~IIIC, Fig.~S13\textbf{a,b}. We plot the loss function and accuracy of both training and test data measured at each epoch. After the training process, the trained quantum classifier is able to classify the total training (test) set with accuracy $0.98$ ($0.99$).

\vspace{.3cm}
\noindent\textbf{Quantum classifiers with quantum data}

\noindent On our device, the frequency of each qubit and the coupling strength between neighboring qubits are programmable with high flexibility, such that we can synthesize the Aubry-Andr\'{e} Hamiltonian (Eq.~\ref{AAH}) and modulate its relevant coefficients such as the coupling strength $g$ and the on-site disorder $V_k$ in arbitrary manners. Experimentally we fix $g$ by setting the coupler frequencies and apply desired flux bias to each qubit to vary $V_k$ as a consine function over $k$.

With the experimental settings introduced above, we construct the training (test) set with $500$ ($100$) quantum states, where half of the states come from the localized phase and the remaining half from the delocalized phase. 
The classifier is composed of five blocks and contains a total number of $150$ training parameters encoded in the single-qubit rotation angles (see Supplementary Sec.~IIIB, Fig.~S9 for the full circuit of the classifier). 
The training parameters are divided into $10$ groups with each group containing $15$ parameters and trained sequentially at each epoch. 
For each group, we randomly select $20$ ($50$) samples to form the training (test) set. 

\vspace{.3cm}
\noindent\textbf{Adversarial training}

\noindent 
The adversarial examples aim to lead the well-trained quantum classifier to make incorrect predictions.
In general, these adversarial examples are generated by adding carefully-designed but imperceptible perturbations to the original samples.
To generate these adversarial perturbations in our work, we have designed several untargeted white-box attack strategies for both the classical and the quantum data, which are described in detail in the Supplementary Sec.~IB and Sec.~IIB.
Essentially, the perturbation is designed to maximize the loss function, which is in line with maximizing the distance between the model's output and the correct label, i.e., effectively deceiving the classifier to make incorrect classifications.
In our work, we utilize this idea and apply gradient ascent methods to generate adversarial perturbations assisted by the Adam optimizer, and the attacking strategies for the classical dataset and the quantum dataset are presented as follows.

First, we consider the case of classical data.
For each sample in the training (test) set with size $500$ ($100$),
we numerically generate a corresponding adversarial example on a classical computer aiming to lead the well-trained classifier to make an incorrect prediction.
We calculate the gradients of the loss function with respect to the input sample and use gradient ascent to maximize the loss function.
For concreteness, two strategies are applied to generate two types of adversarial examples, namely, type-$1$ examples and type-$2$ examples
(see Supplementary Sec.~IB for detailed algorithms).
These generated adversarial examples are then processed by the quantum classifier.
As shown in Fig.~\ref{fig2}\textbf{e} and \ref{fig2}\textbf{f}, we experimentally verify the effectiveness of these adversarial examples, where the quantum classifier tends to assign incorrect labels to them.
Moreover, we provide supplementary experimental demonstrations of adversarial examples with the MNIST handwritten digit dataset in Supplementary Sec.~IIIC, Fig.~S13\textbf{c}, from which we can see that the slightly-perturbed handwritten digits successfully deceive the quantum classifier.
We mention that this procedure requires high-quality superconducting quantum processors, so that the adversarial examples generated by a classical computer can still deceive the quantum classifier, despite the inevitable experimental noises. 

Second, 
to generate the adversarial examples for quantum data, we add local perturbation, which is parameterized by three single-qubit gates, i.e., $R_x(\delta_1)R_z(\delta_2)R_x(\delta_3)$ with $\delta_i \in [-0.5,0.5]$, to each qubit before tuning the system to evolve under the Aubry-Andr\'{e} Hamiltonian.
These perturbations are optimized experimentally to maximize the loss function, i.e., to lead the quantum classifier to make incorrect predictions.
To ensure that the locally-perturbed states maintain the original states' property (localized or thermal),
we compare the states before and after adding adversarial perturbations experimentally (Fig.~\ref{fig3}\textbf{d}).
For more information about generating adversarial examples for both classical data and quantum data, we provide the detailed algorithms in Supplementary Sec.~IB and Sec.~IIB.

Now, we introduce the settings for the adversarial training of quantum classifiers.
The basic idea is to mix the adversarial samples and the original samples to construct new training and test sets.
We start the training by re-initializing the $260$ trainable parameters with random values.
At each training epoch, we randomly select $10$ samples from original data set and $10$ from the adversarial data set to form a training batch. 
The learning rate and the optimization strategies remain the same as those in the original training procedure.
After the re-training process,
the loss function and accuracy for both the original and adversarial samples measured at each training step are shown in Fig.~\ref{fig4}\textbf{a} with a specific example shown in Fig.~\ref{fig4}\textbf{b}.
And it turns out that the re-trained classifier is able to identify both the legitimate samples and the adversarial ones with high accuracy, and thus has obtained the immunity against certain adversarial attacks.
Similarly, the same adversarial training has been successfully implemented with the MNIST handwritten digit dataset, with the obtained experimental results shown in Supplementary Sec.~IIIC, Fig.~S13\textbf{d}.

\vspace{.3cm}

\vspace{.6cm}
\noindent\textbf{\large{}Data availability}
The data presented in the figures and that support the other findings of this study are available upon reasonable request from the corresponding authors. 

\vspace{.6cm}
\noindent\textbf{\large{}Code availability}
The data analysis and numerical simulation codes are available from the corresponding authors on reasonable request.

\vspace{.5cm}
\noindent\textbf{Acknowledgement} We thank L.-M.~Duan and Sirui Lu for helpful discussions, and Vedran Dunjko in particular for his valuable feedback from reading the first version of this paper. The device was fabricated at the Micro-Nano Fabrication Center of Zhejiang University. We acknowledge the support of the National Natural Science Foundation of China (Grants No. 92065204, No. U20A2076, No. 11725419, No. 12174342, and 12075128), the National Basic Research Program of China (Grants No. 2017YFA0304300), the Zhejiang Province Key Research and Development Program (Grant No. 2020C01019), the Key-Area Research and Development Program of Guangdong Province (Grant No. 2020B0303030001), and the Fundamental Research Funds for the Zhejiang Provincial Universities (Grant No. 2021XZZX003). D.-L. D. also acknowledges additional support from the Shanghai Qi Zhi Institute.

\vspace{.3cm}
\noindent\textbf{Author contributions}  
D.-L.D.~conceived the experiment; W.R.~and S.X.~carried out the experiments supervised by C.S.~and H.W.; H.L.~fabricated the device supervised by H.W.; W.L.~and W.J.~performed the numerical simulations supervised by D.-L.D. All authors contributed to the analysis of data, the discussions of the results and the writing of the manuscript.

\vspace{.3cm}
\noindent\textbf{Competing interests}  All authors declare no competing interests.

\vspace{.3cm}
\noindent{* These authors contributed equally to this work.\\
$^\dagger$ chaosong@zju.edu.cn\\
$^\S$ dldeng@tsinghua.edu.cn \\
$^\ddagger$ hhwang@zju.edu.cn}

\bibliography{QMLBib,Dengbib,QAMLsupp}

\clearpage
\newpage 
\onecolumngrid
\setcounter{section}{0}
\setcounter{equation}{0}
\setcounter{figure}{0}
\setcounter{table}{0}
\setcounter{page}{1}
\makeatletter
\renewcommand\thefigure{S\arabic{figure}}
\renewcommand\thetable{S\arabic{table}}
\renewcommand\theequation{S\arabic{equation}}

\begin{center} 
	{\large \bf Supplementary Information: Experimental quantum adversarial learning with programmable superconducting qubits}
\end{center}

\setcounter{figure}{0}
\setcounter{table}{0}
\renewcommand\thefigure{S\arabic{figure}}
\renewcommand\thetable{S\arabic{table}}
\maketitle
\tableofcontents

\section{Theoretical details for quantum neural networks handling classical data}
\label{app:theory1}

\subsection{Quantum neural network classifiers}
\label{app:classifiers}
With the recent development in quantum machine learning \cite{Biamonte2017Quantum,Dunjko2018Machine,Sarma2019Machine}, some advanced quantum machine learning algorithms may bring near-term applications. 
In the era of noisy intermediate-scale quantum (NISQ) devices \cite{Preskill2018Quantum}, variational quantum algorithms have been developed tremendously \cite{Cerezo2021Variational}, among which quantum neural network (QNN) classifiers have drawn a wide range of interest over the recent years \cite{Li2022Recent}.
In this subsection, we will introduce the basic structures and optimization strategies for QNN classifiers.
To experimentally demonstrate QNN classifiers with high-dimensional datasets, we introduce an ``interleaved'' QNN architecture which has the expressive power to handle the classification of real-life images up to $256$-dimensional, followed by numerical benchmarks for exhibiting the better classification performance than the ``encoding first'' QNN architecture.

\subsubsection{Basic structures}
\label{app:structure}
Quantum neural networks are usually considered as the quantum analog of classical neural networks, whose structures can be represented by parameterized quantum circuits.
For the basic building blocks of QNN circuits, popular choices include single-qubit rotation gates and two-qubit controlled gates:

$$
\begin{quantikz}
& \gate{R_x(\theta)} & \qw
\end{quantikz}
= e^{-i \frac{\theta}{2} \hat{\sigma}_x}
\begin{quantikz}
& \gate{R_y(\theta)} & \qw
\end{quantikz}
= e^{-i \frac{\theta}{2} \hat{\sigma}_y}
\begin{quantikz}
& \gate{R_z(\theta)} & \qw
\end{quantikz}
= e^{-i \frac{\theta}{2} \hat{\sigma}_z}
$$

$$
\begin{quantikz}
& \ctrl{1} & \qw \\
& \targ{} & \qw
\end{quantikz}
=\begin{pmatrix}
1 & 0 & 0 & 0\\
0 & 1 & 0 & 0\\
0 & 0 & 0 & 1\\
0 & 0 & 1 & 0\\
\end{pmatrix}
\begin{quantikz}
& \ctrl{1} & \qw \\
& \gate{Z} & \qw
\end{quantikz}
=\begin{pmatrix}
1 & 0 & 0 & 0\\
0 & 1 & 0 & 0\\
0 & 0 & 1 & 0\\
0 & 0 & 0 & -1\\
\end{pmatrix}
$$
In practice, there are many other choices for experimental demonstrations such as the Controlled-SWAP gate and the iSWAP gate. Which one to choose should take into account the platform and the detailed task information. In addition to the digital components listed above, the evolution of a global Hamiltonian can be utilized as an analog block. In our work, we mainly use single-qubit gates ($R_x(\theta)$, $R_y(\theta)$, and $R_z(\theta)$) and Controlled-NOT gates as the building blocks for experimental demonstrations and numerical benchmarks.

With a QNN structure constructed using the chosen building blocks, it can be utilized to handle some optimization-based tasks, where the rotation angles in the single-qubits gates can be used as variational parameters.
For classification tasks, we need to encode the input data into the QNN classifier.
If the data directly comes from a quantum process, we can assume that it is already encoded into the input quantum state and can be fed into a quantum classifier directly.
However, if the data is from a classically-stored dataset, which is often seen as a suitable case for implementations on NISQ devices, we need to encode it into the QNN circuit.
In this situation, one method is to encode the data into the rotation angles of the single-qubits gates similar to the encoding of variational parameters.

\begin{figure*}[t]
	\includegraphics[width=0.7\textwidth]{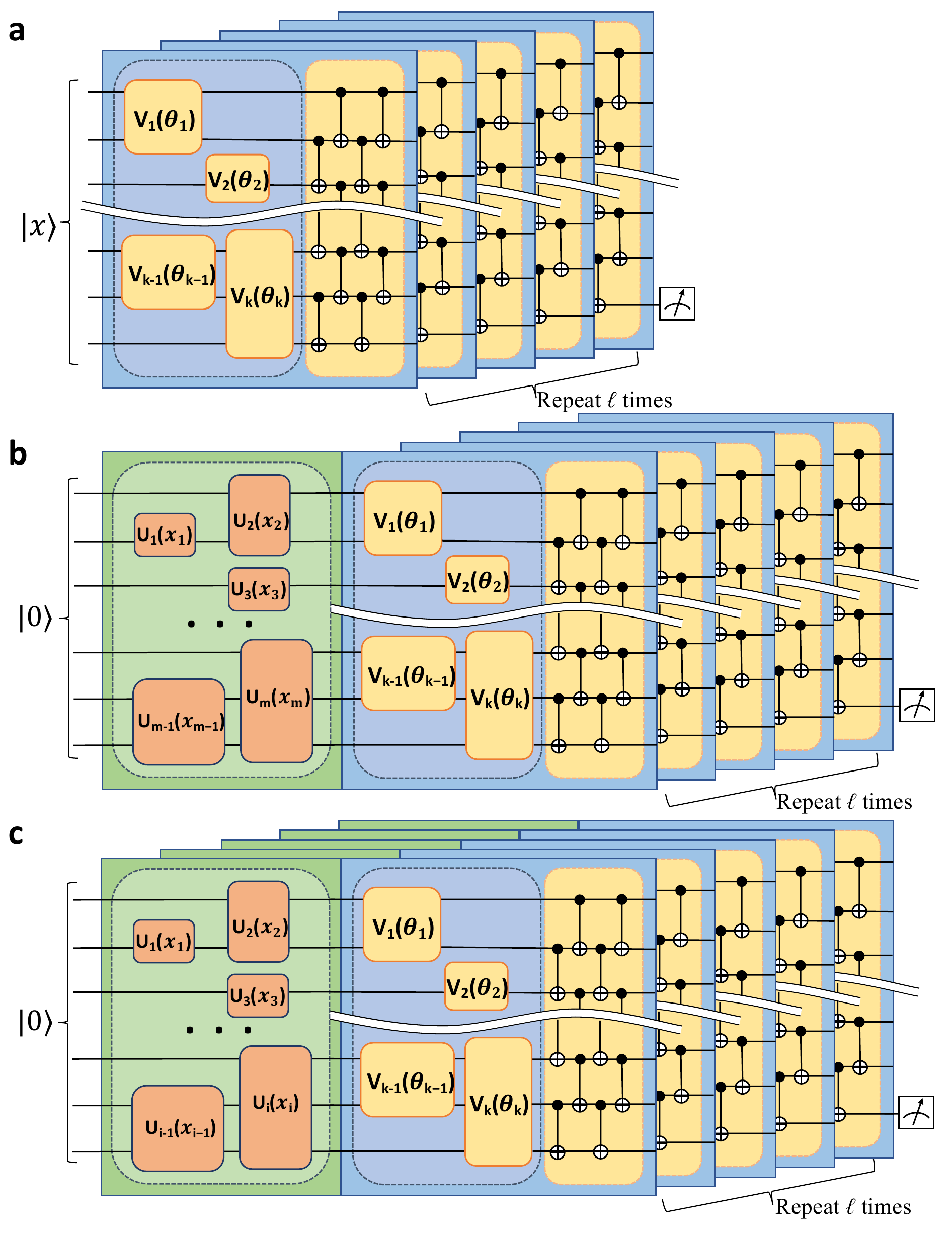}
	\caption{\textbf{Schematics of three encoding strategies for QNN classifiers.} \textbf{a}, The amplitude-encoding QNN structure, where we assume the input state already encodes the data, followed by a variational QNN circuit; \textbf{b}, The ``encoding first'' block-encoding QNN structure, where the first part of the QNN is used to encode the data, followed by a variational part to be trained; \textbf{c}, The ``interleaved'' block-encoding QNN structure, where the data-encoding blocks and variational blocks are interleaved.}
	\label{encoding}
\end{figure*}

As shown in Fig.~\ref{encoding}, we present three QNN structures:
(1) The amplitude-encoding QNN structure, where we assume the input state contains the data information; 
(2) The ``encoding first'' block-encoding QNN structure, where the first part of the QNN is used to encode the data, followed by a variational part to be trained;
(3) The ``interleaved'' block-encoding QNN structure, where the data-encoding blocks and variational blocks are interleaved.
The third structure is utilized in our experiments, and in Sec. \ref{app:benchmark}, we will numerically benchmark the performances for the second and third one.

\subsubsection{Optimization strategies}
\label{app:opt}

With the QNN structure discussed above, our goal is to train a QNN classifier which is able to learn the patterns from the training data and has decent generalization performance on the test set.
Thus, first we need to formalize the task to be an optimization problem.
For both amplitude encoding and block encoding schemes, the output is chosen as an expectation value of some observables, according to which the classification decisions are made.
For example, when we adapt the QNN classifier to recognize different medical images labeled ``benign'' and ``malicious'', we can choose the expectation value of the $Z$-basis measurement on the last qubit.
If the label of the input data is ``benign'', our goal is to train the QNN classifier to maximize the expectation value, i.e., maximize $P(\ket{0})$.
If the label is ``malicious'', then the goal is to minimize the expectation value, i.e., maximize $P(\ket{1})$.
After the training phase, the predictions of unseen samples are made according to $\operatorname{argmax}\{P(\ket{0}),P(\ket{1})\}$.
The basic settings for the prediction phase are listed below:

\begin{itemize}
	\item In our work, we mainly consider binary classification tasks. Given an input $\mathbf{x}$ and trainable parameters $\boldsymbol{\theta}$: (1) For block encoding schemes, the output state will be $\ket{\Psi} = U_{\mathbf{x},\boldsymbol{\theta}}\ket{00...0}$; (2) For amplitude encoding schemes, the output state will be $\ket{\Psi} = U_{\boldsymbol{\theta}}\ket{x}$. We define the observables of the binary measurements on the Pauli $Z$-basis as the projectors $\mathcal{O}_{k}^{+}$ and $\mathcal{O}_{k}^{-}$ corresponding to spins $+1$ and $-1$, respectively, where $k$ denotes the index of the qubit on which we apply our measurements.
	\item It is obvious that $\bra{\Psi}(\mathcal{O}_{k}^{+}+\mathcal{O}_{k}^{-})\ket{\Psi} = 1$. Now we define the probability of assigning $\ket{\Psi}$ to class $1$ as $P_1(\ket{\Psi}) = \bra{\Psi}\mathcal{O}_{k}^{+}\ket{\Psi}$, and to class $2$ as $P_2(\ket{\Psi}) = \bra{\Psi}\mathcal{O}_{k}^{-}\ket{\Psi}$. Given a new input $\ket{\Psi_i}$, it will be assigned to class $1$ if $P_1(\ket{\Psi_i}) > P_2(\ket{\Psi_i})$ and vice versa. With a trained model, the predictions are expected to agree with the true labels.
\end{itemize}

With these definitions and the training goals, we now discuss how to achieve these goals through an optimization procedure.
In general deep supervised learning tasks, we need a loss function to measure the distance between the current predictions and target predictions.
Easy-understand examples include the mean square error (MSE):
\begin{eqnarray}
\mathcal{L}_{MSE}\left(h\left(\mathbf{x} ; \boldsymbol{\theta}\right), \mathbf{a}\right)=\sum_{k} \left(a_{k} - g_{k}\right)^2,
\end{eqnarray}
where $\mathbf{a}\equiv (a_1, \cdots, a_m)$  denotes the label of the input $\mathbf{x}$ in the form of one-hot encoding,
$h$ denotes the hypothesis function determined by the QNN (with parameters collectively denoted by $\boldsymbol{\theta}$),
and $\mathbf{g}\equiv (g_1,\cdots, g_m)=\text{diag} (\rho_{\text{out}})$ presents the probabilities of the output categories in the standard basis with $\rho_{\text{out}}$ denoting the output state \cite{Lu2020Quantum}.
More specifically, for amplitude-encoding schemes, $g_k = h_k\left(|\psi_{x}\rangle ; \boldsymbol{\theta}\right) = \bra{x} U^{\dagger}_{\boldsymbol{\theta}} \mathcal{O}_k U_{\boldsymbol{\theta}} \ket{x}$; meanwhile, for block-encoding schemes, $g_k = h_k\left(\mathbf{x} ; \boldsymbol{\theta}\right) = \bra{0} U^{\dagger}_{\boldsymbol{\theta},\mathbf{x}} \mathcal{O}_k U_{\boldsymbol{\theta},\mathbf{x}}\ket{0}$.
The MSE clearly exhibits the goal of the training, i.e., minimizing the difference between the target predictions and QNN's outputs. 

In our work, we choose the cross entropy as the loss function:
\begin{eqnarray}
\mathcal{L}_{CE}\left(h\left(\mathbf{x} ; \boldsymbol{\theta}\right), \mathbf{a}\right)=-\sum_{k} a_{k} \log g_{k},
\end{eqnarray}
and for binary classifications, it can be written as
\begin{eqnarray}
	\mathcal{L}_{CE}\left(h\left(\mathbf{x} ; \boldsymbol{\theta}\right), \mathbf{a}\right)=-a_{1} \log g_{1}-a_{2} \log g_{2}.
\end{eqnarray}

If a new sample belongs to class $1$, i.e., $a_1 = 1$, $a_2 = 0$.
Then the loss function can be further reduced to
\begin{eqnarray}
	\mathcal{L}_{CE}\left(h\left(\mathbf{x} ; \boldsymbol{\theta}\right), \mathbf{a}\right)=-a_{1} \log g_{1}.
\end{eqnarray}

\begin{figure}
\begin{algorithm}[H]
\caption{Quantum neural network classifier for classifying the medical data}
\label{Algo_Medical}
\begin{algorithmic}[1]
\REQUIRE The model $h$ with parameters $\boldsymbol{\theta}$, the loss function $\mathcal{L}$, the number of samples $n$, the training set $\{(\mathbf{x}_m, \mathbf{a}_m)\}_{m=1}^n$, the batch size $n_b$, the number of iterations $T$, the learning rate $\epsilon$, and the Adam optimizer $f_{\text{Adam}}$
\ENSURE The trained model
\STATE Initialization: generate random initial parameters for $\boldsymbol{\theta}$
\FOR{ $i \in [T]$}
    \STATE Divide the $260$ variational parameters into $10$ parameter-batches $\{b_1,b_2,...,b_{10}\}$, with each parameter-batch denoting the parameters encoded on the same qubit (i.e., the same row in the QNN circuit)
    \FOR{ $j \in [10]$}
        \STATE Randomly choose $n_b$ samples $\{\mathbf{x}_{\text{(i,j,1)}},\mathbf{x}_{\text{(i,j,2)}},...,\mathbf{x}_{\mathrm{(i,j,n_b)}}\}$ among the $n$ samples in the training set
        \STATE Calculate the gradients for parameter-batch $b_j$ in experiments using the ``parameter shift rule'', and take the average value over the training batch $\mathbf{G}\leftarrow\frac{1}{n_b}\Sigma_{k=1}^{n_b}\nabla \mathcal{L}(h(\mathbf{x}_{\text{(i,j,k)}};b_j),\mathbf{a}_{\text{(i,j,k)}})$
        \STATE Updates: $b_j \leftarrow f_{\text{Adam}}(b_j,\epsilon,\mathbf{G})$
    \ENDFOR
\ENDFOR
\STATE Output the trained model
\end{algorithmic}
\end{algorithm}
\end{figure}

\begin{figure}
\begin{algorithm}[H]
\caption{Quantum neural network classifier for classifying the MNIST data}
\label{Algo_MNIST}
\begin{algorithmic}[1]
\REQUIRE The model $h$ with parameters $\boldsymbol{\theta}$, the loss function $\mathcal{L}$, the number of samples $n$, the training set $\{(\mathbf{x}_m, \mathbf{a}_m)\}_{m=1}^n$, the batch size $n_b$, the number of iterations $T$, the learning rate $\epsilon$, and the Adam optimizer $f_{\text{Adam}}$
\ENSURE The trained model
\STATE Initialization: generate random initial parameters for $\boldsymbol{\theta}$
\FOR{ $i \in [T]$}
    \STATE Randomly choose $n_b$ samples $\{\mathbf{x}_{\text{(i,1)}},\mathbf{x}_{\text{(i,2)}},...,\mathbf{x}_{\mathrm{(i,n_b)}}\}$ among the $n$ samples in the training set
    \STATE Choose $50$ variational parameters among the $260$ available ones, which lie at the $3$rd, $6$th, $11$th, $17$th, $23$rd columns of the QNN circuit
    \STATE Calculate the gradients in experiments using the ``parameter shift rule'', and take the average value over the training batch $\mathbf{G}\leftarrow\frac{1}{n_b}\Sigma_{k=1}^{n_b}\nabla \mathcal{L}(h(\mathbf{x}_{\text{(i,k)}};\boldsymbol{\theta}),\mathbf{a}_{\text{(i,k)}})$
    \STATE Updates: $\boldsymbol{\theta} \leftarrow f_{\text{Adam}}(\boldsymbol{\theta},\epsilon,\mathbf{G})$
\ENDFOR
\STATE Output the trained model
\end{algorithmic}
\end{algorithm}
\end{figure}

To minimize the loss function, we adapt the gradient descent method.
Here, computing the derivatives of $\mathcal{L}$ with respect to the circuit parameters can be transformed into computing the derivatives of some expectation values with respect to these circuit parameters according to the chain rule.
In our case, it can be formally expressed as
\begin{eqnarray}
\frac{\partial \mathcal{L}_{CE}\left(h\left(\mathbf{x} ; \boldsymbol{\theta}\right), \mathbf{a}\right)}{\partial \theta} = -\sum_{k} \frac{a_{k}}{g_{k}} \frac{\partial g_{k}}{\partial \theta}.
\end{eqnarray}
The next step that computes the derivatives of $g_k$ with respect to the circuit parameters can be accomplished with the ``parameter shift rule'', since $g_k$ can be regarded as an expectation value of an observable which we denote as $B_k$ here
\cite{Mitarai2018Quantum,Li2017Hybrid,Schuld2019Evaluating}.
This rule states that if a gate with parameter $\theta$ is in the form $\mathcal{G}(\theta)=e^{-i \frac{\theta}{2} P_n}$ with $P_n$ being an $n$-qubit Pauli string, 
the derivative can be evaluated by:
\begin{eqnarray}
\frac{\partial g_k}{\partial \theta}=\frac{\partial \langle B_k\rangle}{\partial \theta}=\frac{\langle B_k\rangle^{+}-\langle B_k\rangle^{-}}{2},
\end{eqnarray}
where $\langle B_k\rangle^{\pm}$ denotes the expectation values of $B_k$ (i.e., $\bra{\Psi}\mathcal{O}_k\ket{\Psi}$) with the parameter $\theta$ being $\theta \pm \frac{\pi}{2}$.
Thus, since the parameters in our case are all encoded in the angles of single-qubit Pauli-rotation gates,
we can optimize the QNN classifier with gradients obtained from measurements.
Compared with finite difference methods such as $\frac{\partial\langle B\rangle}{\partial \theta} \approx \frac{\langle B\rangle_{\theta+\frac{\Delta \theta}{2}}-\langle B\rangle_{\theta-\frac{\Delta \theta}{2}}}{\Delta \theta}$, 
the ``parameter shift rule'' provides exact gradients without discretization error,
and it is convenient to be implemented on the near-term quantum devices.

Next, we can update the trainable parameters $\boldsymbol{\theta}$ by gradient descent:
\begin{eqnarray}
\boldsymbol{\theta}_{t+1} = \boldsymbol{\theta}_{t} - \epsilon \cdot \nabla \mathcal{L}\left(\boldsymbol{\theta}_{t}\right),
\end{eqnarray}
where $\boldsymbol{\theta}_{t} $ denotes collectively the parameters at the $t$-th step, $\epsilon$ is the learning rate.
In practice, we take the Adam optimizer for higher training performance \cite{Kingma2014Adam}.

\subsubsection{Algorithms and benchmarks}
\label{app:benchmark}

\begin{figure*}[t]
	\includegraphics[width=1\textwidth]{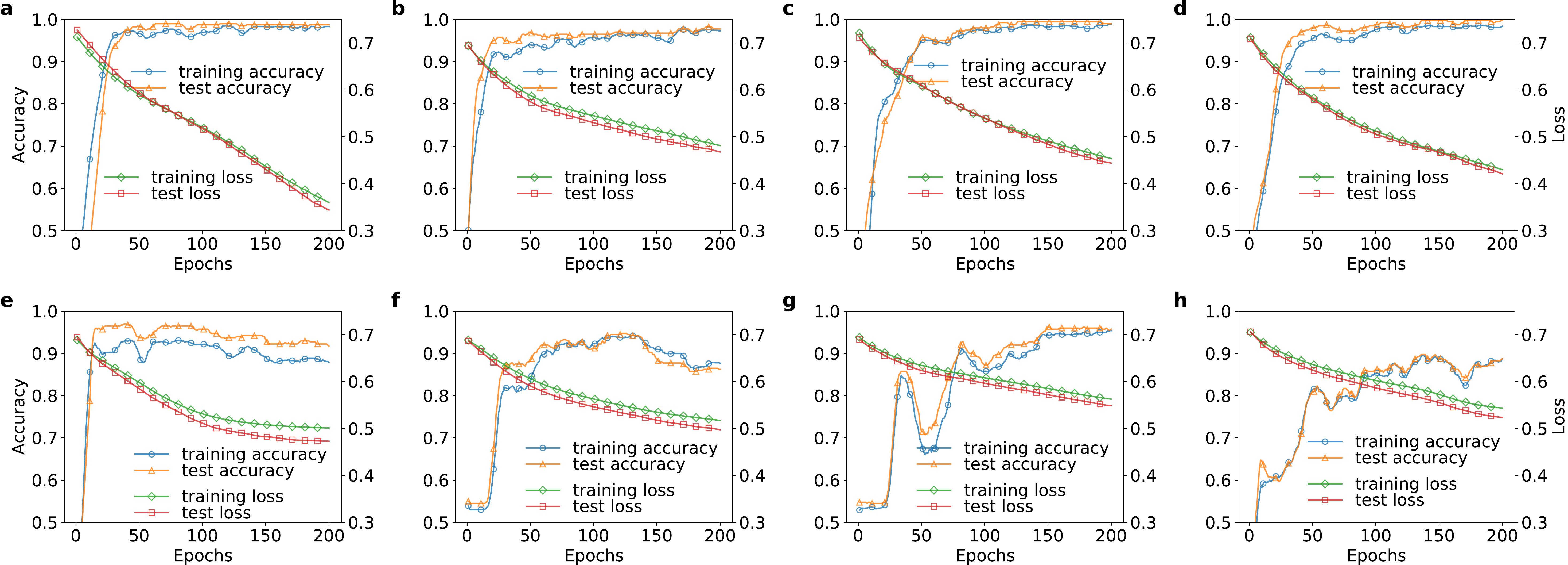}
	\caption{\textbf{Benchmarks for ``interleaved'' block-encoding QNN structures and ``encoding first'' block-encoding QNN structures with the MNIST dataset.} \textbf{a-d}, For each figure, we assign random initial parameters to an ``interleaved'' block-encoding QNN classifier and the accuracy and loss curves are separately shown. \textbf{e-h}, For each figure, we assign random initial parameters to an ``encoding first'' block-encoding QNN classifier and the accuracy and loss curves are separately shown.}
	\label{benchmark_mnist}
\end{figure*}
\begin{figure*}[t]
	\includegraphics[width=0.8\textwidth]{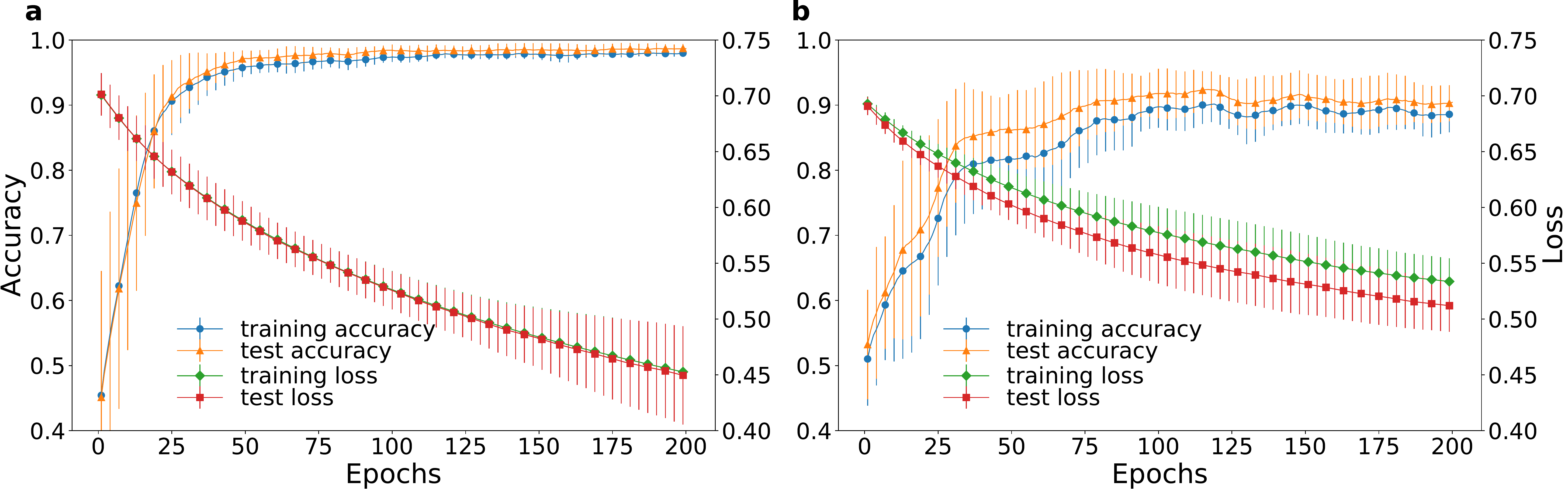}
	\caption{\textbf{Benchmarks for ``interleaved'' block-encoding QNN structures and ``encoding first'' block-encoding QNN structures with the MNIST dataset.} \textbf{a}, We assign random initial parameters to an ``interleaved'' block-encoding QNN classifier for ten times, and exhibit the accuracy and loss curves averaged over them as well as the standard bias. \textbf{b}, For each figure, we assign random initial parameters to an ``encoding first'' block-encoding QNN classifier for ten times, and exhibit the accuracy and loss curves averaged over them as well as the standard bias.}
	\label{benchmark_mnist_std}
\end{figure*}
\begin{figure*}[t]
	\includegraphics[width=1\textwidth]{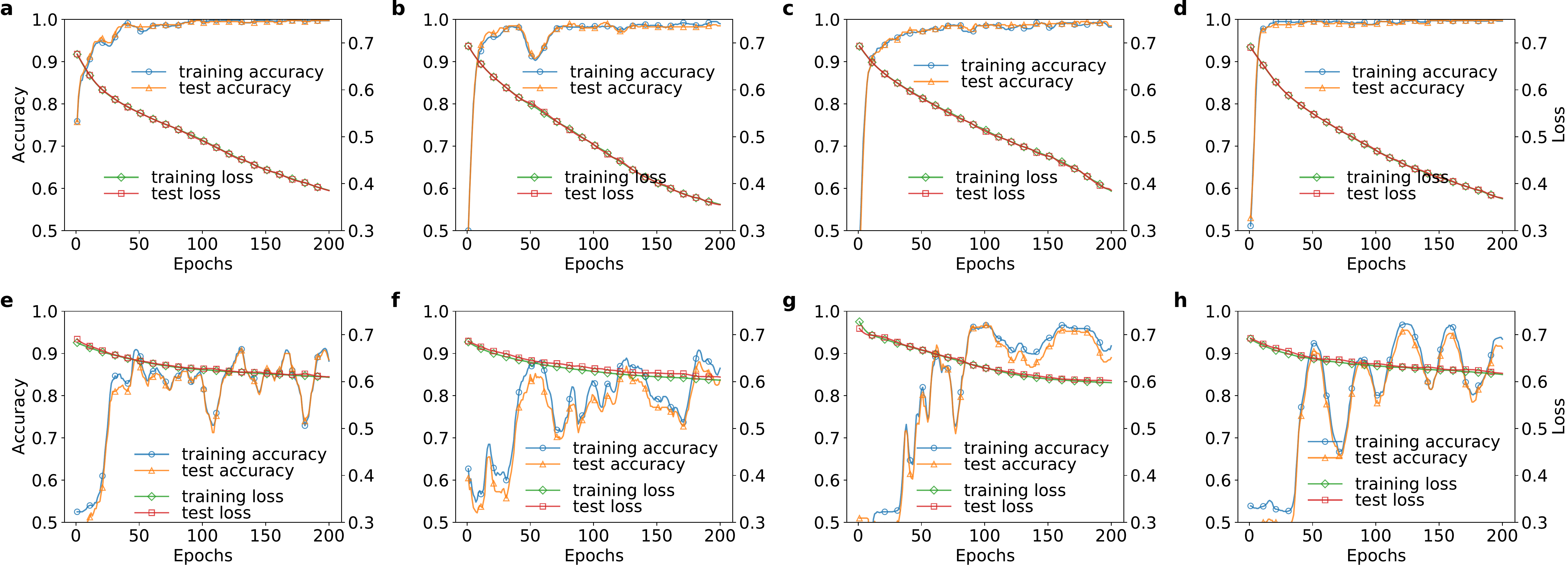}
	\caption{\textbf{Benchmarks for ``interleaved'' block-encoding QNN structures and ``encoding first'' block-encoding QNN structures with the FashionMNIST dataset.} \textbf{a-d}, For each figure, we assign random initial parameters to an ``interleaved'' block-encoding QNN classifier and the accuracy and loss curves are separately shown. \textbf{e-h}, For each figure, we assign random initial parameters to an ``encoding first'' block-encoding QNN classifier and the accuracy and loss curves are separately shown.}
	\label{benchmark_fashionmnist}
\end{figure*}
\begin{figure*}[t]
	\includegraphics[width=0.8\textwidth]{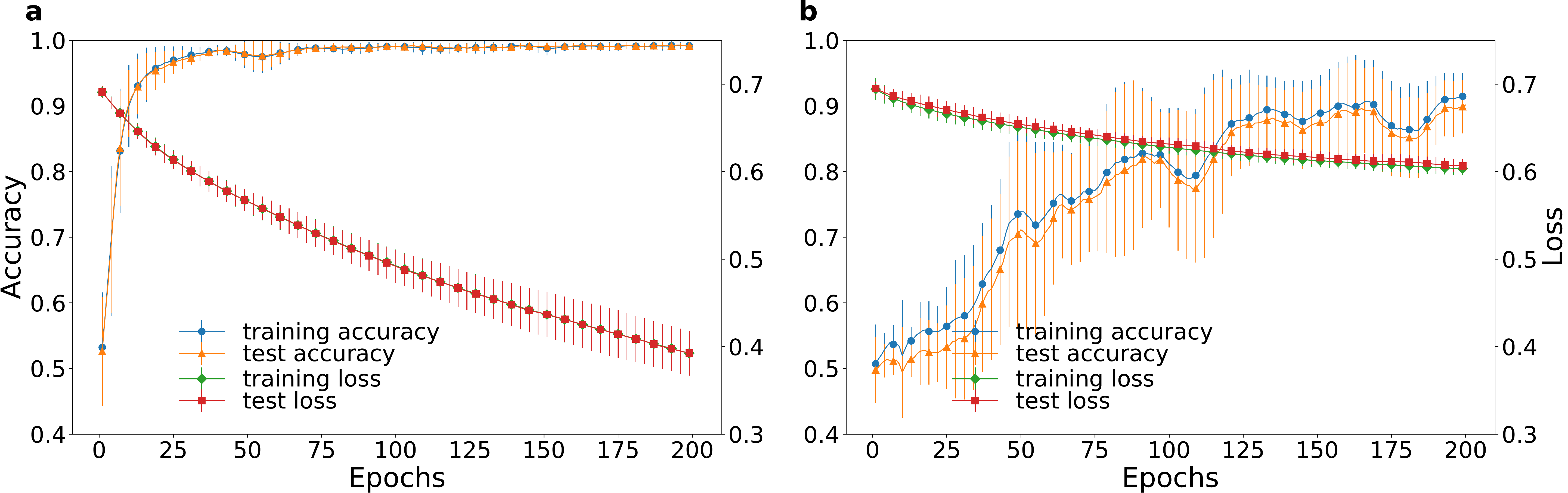}
	\caption{\textbf{Benchmarks for ``interleaved'' block-encoding QNN structures and ``encoding first'' block-encoding QNN structures with the FashionMNIST dataset.} \textbf{a}, For each figure, we assign random initial parameters to an ``interleaved'' block-encoding QNN classifier for ten times, and exhibit the accuracy and loss curves averaged over them as well as the standard bias. \textbf{b}, For each figure, we assign random initial parameters to an ``encoding first'' block-encoding QNN classifier for ten times, and exhibit the accuracy and loss curves averaged over them as well as the standard bias.}
	\label{benchmark_fashionmnist_std}
\end{figure*}

In our experiments, we have designed a QNN structure exhibited in the main text.
For the two training tasks, we deploy two slightly different algorithms for the medical dataset and the MNIST handwritten digit dataset, respectively.
The pseudocode for training on the medical dataset is shown in Algorithm \ref{Algo_Medical}, where we use $260$ parameters (divided into $10$ batches according to the $10$ qubit indexes, each with $26$ parameters) for updating.
The pseudocode for training on the MNIST handwritten digit dataset is show in Algorithm \ref{Algo_MNIST}, where we limit the number of variational parameters to $50$ to reduce the time cost.
In both the two algorithms, the superconducting platform provides high circuit fidelity to calculate the gradients and to optimize the QNN circuit, providing a foundation for further works including demonstrating adversarial attacks and adversarial training.

For block-encoding schemes, we have proposed an ``interleaved'' block-encoding QNN structure to encode the classical data into the QNN circuit.
The reason why we choose this structure over the ``encoding first'' block-encoding QNN structure is explained as follows:

The ``encoding first'' block-encoding QNN structure corresponds to the unitary $U_{\mathbf{x},\boldsymbol{\theta}} = W_{\boldsymbol{\theta}}V_{\mathbf{x}}$,
with $W_{\boldsymbol{\theta}}$ and $V_{\mathbf{x}}$ being the variational part and the encoding part, respectively.
Without loss of generality, we assume the initial input state is $\ket{0}$.
The expectation value of the output state on observable $\mathcal{O}$
is
$g = h\left(|0\rangle, \mathbf{x} ; \boldsymbol{\theta}\right) = \bra{0} V_{\mathbf{x}}^{\dagger}W_{\boldsymbol{\theta}}^{\dagger} \mathcal{O} W_{\boldsymbol{\theta}}V_{\mathbf{x}}\ket{0}$.
Given a threshold $b$, the classification decision is made according to whether $g>b$ or $g<b$.
From the view of a support vector machine (SVM), the ``encoding first'' QNN model can be described by a SVM with a kernel matrix $\mathcal{K}$ where $\mathcal{K}_{ij} = |\bra{0} V_{\mathbf{x}_i}^{\dagger}V_{\mathbf{x}_j}\ket{0}|^2$
\cite{Havlicek2019Supervised,Schuld2021Supervised,Huang2021Power}.
In Ref.~\cite{Havlicek2019Supervised}, the authors pointed out that if the inner product of these states can be evaluated efficiently on a classical computer, then the quantum model can not provide an advantage over classical SVMs.
For our purpose, we aim to design a QNN classifier with high expressive power.
From the above discussion, we see that in the ``encoding first'' case, once the data is encoded,
the performance of the QNN classifier is already upper-bounded by a classical SVM whose kernel matrix is fixed and difficult to predefine.
In practice, this QNN classifier may perform worse than the corresponding classical SVM since the ``linear coefficients'' contained in $W_{\boldsymbol{\theta}}^{\dagger} \mathcal{O} W_{\boldsymbol{\theta}}$ are constrained by the $W_{\boldsymbol{\theta}}^{\dagger} \mathcal{O} W_{\boldsymbol{\theta}}$'s Hermitian property.
On the other hand, with an ``interleaved'' block-encoding QNN structure, we can decompose the unitary $U^{\prime}_{\mathbf{x},\boldsymbol{\theta}} = W^{\prime}_{\boldsymbol{\theta}}V^{\prime}_{\mathbf{x},\boldsymbol{\theta}}$ for clarity.
Intuitively, in this way not only the ``linear coefficients'' contained in $W_{\boldsymbol{\theta}}^{\prime \dagger} \mathcal{O} W_{\boldsymbol{\theta}}^{\prime}$ can be adjusted during the training process, the kernel $\mathcal{K}^{\prime}$ where $\mathcal{K}_{i j}^{\prime}=\left|\left\langle 0\left|V_{\mathbf{x}_{i}, \boldsymbol{\theta}^{\dagger}}^{\dagger} V_{\mathbf{x}_{j}, \boldsymbol{\theta}}^{\prime}\right| 0\right\rangle\right|^{2}$
can also be optimized, adapting the kernel space for better performance.

Here, to illustrate the performances with different encoding strategies, we provide numerical simulations to benchmark the performances of the ``interleaved'' block-encoding QNN structure and the ``encoding first'' block-encoding QNN structure.
We first design a QNN circuit with $540$ parameters,
where we can use $270$ of them to encode the input data and $270$ as variational parameters.
To create an ``interleaved'' block-encoding QNN classifier, we divide the circuit into $9$ blocks with each block encoding $30$ input elements and $30$ variational parameters.
As for a ``encoding first'' block-encoding QNN classifier, 
we simply use the first $270$ parameters in the circuit to encode the input data and leave the rest $270$ ones as variational parameters.
We choose two datasets, the MNIST handwritten digit dataset and the FashionMNIST dataset, of which each has a $1000$-sample training set and a $400$-sample test set. 
The learning rate is set to $0.003$ assisted by the Adam optimizer \cite{Kingma2014Adam}. 
The training procedure is exhibited in Fig.~\ref{benchmark_mnist}, Fig.~\ref{benchmark_mnist_std} and Fig.~\ref{benchmark_fashionmnist}, Fig.~\ref{benchmark_fashionmnist_std}, from which it is obviously shown that the performances of the ``encoding first'' block-encoding QNN classifier are comparably lower than the ``interleaved'' one in practical high-dimensional numerical simulations.
It should be noted that these results does note rule out the practical applications of ``encoding first'' block-encoding strategy, since here our goal is to design effective QNN classification schemes to classify high-dimensional datasets on near-term quantum devices.
The performance of the ``encoding first'' block-encoding strategy is closely related to the kernel matrix that the data-encoding block provides.
With carefully-designed ``encoding first'' QNN structures, this encoding strategy may map a complex dataset to an easy-to-handle kernel space, even with potential quantum advantages \cite{Liu2021Rigorous}.

\subsection{Quantum adversarial machine learning}

\begin{figure}
\begin{algorithm}[H]
\caption{Generating type-$1$ adversarial examples with gradient descent method}
\label{Algo_Adv_global}
\begin{algorithmic}[1]
\REQUIRE The model $h$ with trained parameters $\boldsymbol{\theta}^{*}$, the loss function $\mathcal{L}$, the number of iterations $T$, the learning rate $\epsilon$, the Adam optimizer $f_{\text{Adam}}$, and a legitimate sample $\mathbf{x}$ with label $\mathbf{a}$
\ENSURE The adversarial example $\mathbf{x}^{\mathrm{adv}}$
\STATE Initialization: $\mathbf{x}^{\mathrm{adv}} \leftarrow \mathbf{x}$
\FOR{ $i \in [T]$}
    \STATE Calculate the gradients of the loss function $\mathcal{L}$ with respect to the vector elements of the input sample $\mathbf{G}_{\mathbf{x}}\leftarrow\nabla_{\mathbf{x}} \mathcal{L}(h(\mathbf{x}^{\mathrm{adv}};\boldsymbol{\theta}^{*}),\mathbf{a})$
    \STATE Updates: $\mathbf{x}^{\mathrm{adv}} \leftarrow f_{\text{Adam}}(\mathbf{x}^{\mathrm{adv}},\epsilon,-\mathbf{G}_{\mathbf{x}})$
\ENDFOR
\STATE Output $\mathbf{x}^{\mathrm{adv}}$
\end{algorithmic}
\end{algorithm}
\end{figure}

\begin{figure}
\begin{algorithm}[H]
\caption{Generating type-$2$ adversarial examples with gradient descent method}
\label{Algo_Adv_local}
\begin{algorithmic}[1]
\REQUIRE The model $h$ with trained parameters $\boldsymbol{\theta}^{*}$, the loss function $\mathcal{L}$, the number of iterations $T$, the learning rate $\epsilon$, the Adam optimizer $f_{\text{Adam}}$, and a legitimate sample $\mathbf{x}$ with label $\mathbf{a}$
\ENSURE The adversarial example $\mathbf{x}^{\mathrm{adv}}$
\STATE Initialization: $\mathbf{x}^{\mathrm{adv}} \leftarrow \mathbf{x}$
\FOR{ $i \in [T]$}
    \STATE Calculate the gradients of the loss function $\mathcal{L}$ with respect to the vector elements of the input sample $\mathbf{G}_{\mathbf{x}}\leftarrow\nabla_{\mathbf{x}} \mathcal{L}(h(\mathbf{x}^{\mathrm{adv}};\boldsymbol{\theta}^{*}),\mathbf{a})$
    \STATE Generate a vector $\mathbf{S_{\mathrm{area}}}$ such that $\mathbf{S_{\mathrm{area}}}[i] \leftarrow 1$ if the area of the object in $\mathbf{x}$ covers index $i$ and $\mathbf{S_{\mathrm{area}}}[i] \leftarrow 0$ otherwise
    \STATE Updates: $\mathbf{x}^{\mathrm{adv}} \leftarrow f_{\text{Adam}}(\mathbf{x}^{\mathrm{adv}},\epsilon,-\mathbf{G}_{\mathbf{x}}\cdot\mathbf{S_{\mathrm{area}}})$
\ENDFOR
\STATE Output $\mathbf{x}^{\mathrm{adv}}$
\end{algorithmic}
\end{algorithm}
\end{figure}

Adversarial machine learning studies the vulnerability of machine learning models as well as developing possible defense strategies \cite{Huang2011Adversarial,Goodfellow2014Explaining,Papernot2017Practical,Biggio2018Wild}.
Early studies of adversarial learning date back to the spam detection problem,
where the system tries to identify whether an uploaded email is spam while some malicious parties tries to change some keywords to escape the detection.
With the recent rise in deep learning, some powerful neural networks are able to classify high-dimensional and complex images, bringing various applications to the modern society from face recognition to self-driving cars.
However, the vulnerability of machine learning models poses great challenges for the security and reliability of these applications:
For example, suppose a neural network model is able to identify the type of disease in a medical image from a patient's X-ray examination.
By adding a carefully-designed and imperceptible perturbation to this image,
the model may output a wrong diagnosis,
leading to potential risks to the patient's health.

\begin{figure*}[t]
	\includegraphics[width=0.5\textwidth]{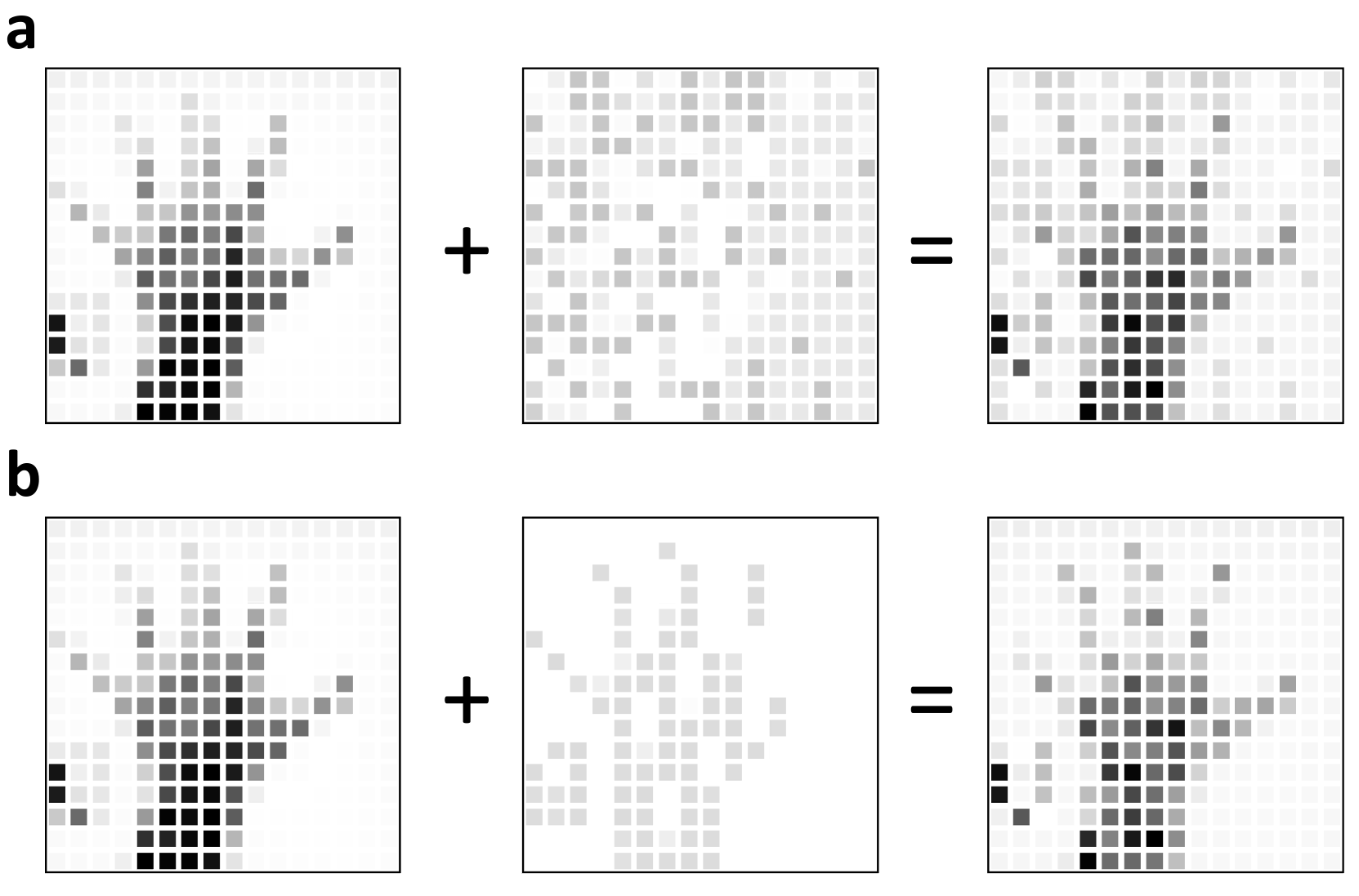}
	\caption{\textbf{Illustration of two types of adversarial attacks used in this work.} \textbf{a}, A type-$1$ adversarial example designed according to Algorithm \ref{Algo_Adv_global}. \textbf{b}, A type-$2$ adversarial example designed according to Algorithm \ref{Algo_Adv_local}.}
	\label{adv_local_global}
\end{figure*}

As mentioned in the main text and in the above subsections, quantum machine learning has achieved dramatic success over the past decade,
with arising works exhibiting the potential quantum advantages over their classical counterparts \cite{Liu2021Rigorous,Abbas2021Power}.
When the quantum machine learning models are able to solve certain practical problems and serve in commercial applications,
the vulnerability of these models should also be granted serious consideration. 
In this subsection, we will briefly introduce adversarial attacks and defense strategies in QNN classifiers as well as the detailed settings of them in our experimental demonstrations.

\subsubsection{Adversarial attacks}

In general, given a QNN model $h\left(\mathbf{x} ; \boldsymbol{\theta}\right)$, we wish to optimize the parameters collectively denoted by $\boldsymbol{\theta}$ such that the loss function $\mathcal{L}\left(h\left(\mathbf{x} ; \boldsymbol{\theta}\right), \mathbf{a}\right)$ is minimized over the training set, i.e., to minimize the distance between the current output and the target output.
By reverse thinking, the idea of adversarial attack is to generate a small perturbation  on the input $\mathbf{x}$ to maximize the distance between the current output and the target output, which can be accomplished by designing a perturbation to maximize the loss function.
As mentioned in the main text, this idea can be formalized as
\begin{eqnarray}
    \boldsymbol{\delta} \equiv \underset{\boldsymbol{\delta^{\prime}} \in \Delta}{\operatorname{argmax}}\; \mathcal{L}\left(h\left(\mathbf{x}+\boldsymbol{\delta^{\prime}} ; \boldsymbol{\theta}^{*}\right), \mathbf{a}\right),
\end{eqnarray}
where $\boldsymbol{\theta}^{*}$ denotes the parameters of a trained model and $\Delta$ restricts the perturbation within a limited region.

In our work, we generate two versions of adversarial examples to experimentally demonstrate the vulnerability of QNN classifiers and the adversarial training.
For the first one, as shown in Algorithm \ref{Algo_Adv_global}, we calculate the gradients of the loss function with respect to the input sample and use gradient ascent to maximize the loss function.
The perturbations are added on the entire image, and we mark them as the type-$1$ adversarial examples.
For the second one, as shown in Algorithm \ref{Algo_Adv_local}, we similarly calculate the gradients of the loss function with respect to the input sample and use gradient ascent methods to maximize the loss function.
The difference is that in this case, we only add perturbation to the area where the object in the image lies in, e.g., for a handwritten digit image,
we only add perturbation on the ``number'' part.
The perturbations are added locally on part of the image, and we mark them as the type-$2$ adversarial examples. In Fig.~\ref{adv_local_global}, we provide an illustrative example to visualize these two strategies. 
We use the type-$1$ adversarial examples to implement the adversarial training.
Moreover, in the main text, we utilize $50$ type-$1$ adversarial examples to exhibit the experimental results of adversarial attacks (Fig.~2f).
The type-$2$ adversarial examples are mainly used for exhibitions in Fig.~2e and Fig.~4b.

\subsubsection{Defense strategies}

As discussed in the main text, the machine learning models are vulnerable to adversarial attacks.
To defend against these potential attacks,
a number of methods have been proposed to enhance the robustness of machine learning models.
Notable examples include adversarial training \cite{Kurakin2017Adversarial}, gradient hiding \cite{Tramer2017Ensemble},
and defensive distillation \cite{Papernot2016Distillation}.
In our experiments, we have demonstrated that QNN classifiers, similar to classical deep learning models, are vulnerable to adversarial attacks.
For the defense strategies, in the main text, we have demonstrated the adversarial training, which turns out to effectively enhance the QNN classifier's robustness.
Here, we provide the detailed algorithms and discussions of the QNN's adversarial training.

\begin{figure}
\begin{algorithm}[H]
\caption{Adversarial training of the medical data}
\label{Algo_Defend_Medical}
\begin{algorithmic}[1]
\REQUIRE The model $h$ with parameters $\boldsymbol{\theta}$, the loss function $\mathcal{L}$, the number of samples $n$, the training set $\{(\mathbf{x}_m, \mathbf{a}_m)\}_{m=1}^n$, the batch size $n_b$, the number of iterations $T$, the learning rate $\epsilon$, and the Adam optimizer $f_{\text{Adam}}$
\ENSURE The trained model
\STATE Initialization: generate random initial parameters for $\boldsymbol{\theta}$
\STATE Generate adversarial examples $\{(\mathbf{x}^{\mathrm{adv}}_m, \mathbf{a}_m)\}_{m=1}^n$ and combine it with the original training set to form an adversarial training set $\mathcal{D}^{\mathrm{adv}} = \{(\mathbf{x}_m, \mathbf{a}_m),(\mathbf{x}^{\mathrm{adv}}_m, \mathbf{a}_m)\}_{m=1}^n$ which has $2n$ samples
\FOR{ $i \in [T]$}
    \STATE Divide the $260$ variational parameters into $10$ parameter-batches $\{b_1,b_2,...,b_{10}\}$, with each parameter-batch denoting the parameters encoded on the same qubit (i.e., the same row in the QNN circuit)
    \FOR{ $j \in [10]$}
        \STATE Randomly choose $n_b$ samples $\{\mathbf{x}_{\text{(i,j,1)}},\mathbf{x}_{\text{(i,j,2)}},...,\mathbf{x}_{\mathrm{(i,j,n_b)}}\}$ among the $2n$ samples in the adversarial training set $\mathcal{D}^{\mathrm{adv}}$
        \STATE Calculate the gradients for parameter-batch $b_j$ in experiments using the ``parameter shift rule'', and take the average value over the training batch $\mathbf{G}\leftarrow\frac{1}{n_b}\Sigma_{k=1}^{n_b}\nabla \mathcal{L}(h(\mathbf{x}_{\text{(i,j,k)}};b_j),\mathbf{a}_{\text{(i,j,k)}})$
        \STATE Updates: $b_j \leftarrow f_{\text{Adam}}(b_j,\epsilon,\mathbf{G})$
    \ENDFOR
\ENDFOR
\STATE Output the trained model
\end{algorithmic}
\end{algorithm}
\end{figure}

\begin{figure}
\begin{algorithm}[H]
\caption{Adversarial training of the MNIST data}
\label{Algo_Defend_MNIST}
\begin{algorithmic}[1]
\REQUIRE The model $h$ with parameters $\boldsymbol{\theta}$, the loss function $\mathcal{L}$, the number of samples $n$, the training set $\{(\mathbf{x}_m, \mathbf{a}_m)\}_{m=1}^n$, the batch size $n_b$, the number of iterations $T$, the learning rate $\epsilon$, and the Adam optimizer $f_{\text{Adam}}$
\ENSURE The trained model
\STATE Initialization: generate random initial parameters for $\boldsymbol{\theta}$
\STATE Generate adversarial examples $\{(\mathbf{x}^{\mathrm{adv}}_m, \mathbf{a}_m)\}_{m=1}^n$ and combine it with the original training set to form an adversarial training set $\mathcal{D}^{\mathrm{adv}} = \{(\mathbf{x}_m, \mathbf{a}_m),(\mathbf{x}^{\mathrm{adv}}_m, \mathbf{a}_m)\}_{m=1}^n$ which has $2n$ samples
\FOR{ $i \in [T]$}
    \STATE Randomly choose $n_b$ samples $\{\mathbf{x}_{\text{(i,1)}},\mathbf{x}_{\text{(i,2)}},...,\mathbf{x}_{\mathrm{(i,n_b)}}\}$ among the $2n$ samples in the adversarial training set $\mathcal{D}^{\mathrm{adv}}$
    \STATE Choose $50$ variational parameters among the $260$ available ones, which lie at the $3$rd, $6$th, $11$th, $17$th, $23$rd columns of the QNN circuit
    \STATE Calculate the gradients in experiments using the ``parameter shift rule'', and take the average value over the training batch $\mathbf{G}\leftarrow\frac{1}{n_b}\Sigma_{k=1}^{n_b}\nabla \mathcal{L}(h(\mathbf{x}_{\text{(i,k)}};\boldsymbol{\theta}),\mathbf{a}_{\text{(i,k)}})$
    \STATE Updates: $\boldsymbol{\theta} \leftarrow f_{\text{Adam}}(\boldsymbol{\theta},\epsilon,\mathbf{G})$
\ENDFOR
\STATE Output the trained model
\end{algorithmic}
\end{algorithm}
\end{figure}

For the adversarial training of both the medical dataset and the MNIST handwritten digit dataset, the basic framework is the same as Algorithm \ref{Algo_Medical} and Algorithm \ref{Algo_MNIST}, respectively.
The difference is that we change the original legitimate training set to a combination of the original training set and the adversarial examples, as shown in Algorithm \ref{Algo_Defend_Medical} and Algorithm \ref{Algo_Defend_MNIST}.
For both two datasets, we generate type-$1$ adversarial examples for adversarial training.
To test the performance, first we can directly check the result in the test set whose adversarial examples follow the same distribution as the adversarial data in the training set.
Moreover,
we generate type-$2$ adversarial examples and check the retrained QNN's performance on these examples.
The latter one is also able to test the transferability of adversarial training from a known adversarial attack to an unknown one in some sense.

\section{Theoretical details for quantum neural networks handling quantum data}
\label{app:theory2}

In the above section, we have already presented the basic concepts for the QNN based supervised learning, quantum adversarial learning, as well as some results from numerical simulations.
In this section, we focus on QNN classifiers handling quantum datasets, where some overlaps with the above section will not be mentioned again.

\subsection{Quantum neural network classifiers}
\label{app:classifiers2}

When handling a quantum dataset, we assume that the input data is already prepared into quantum states.
Thus, unlike the classical data's case, we do not need to encode the data into the QNN circuit, but use an amplitude-encoding QNN structure shown in Fig.~\ref{encoding}\textbf{a} to process the input quantum states.
During the training process, we similarly utilize the parameter shift rule to calculate the gradients and optimize the QNN's parameters according to the strategy shown in Algorithm \ref{Algo_Qdata}.
In the following section of experimental details, we will exhibit the detailed structure handling the quantum dataset sampled from two distinct phases of Aubry-Andr\'{e} model.

\begin{figure}
\begin{algorithm}[H]
\caption{Quantum neural network classifier for classifying the quantum data}
\label{Algo_Qdata}
\begin{algorithmic}[1]
\REQUIRE The model $h$ with parameters $\boldsymbol{\theta}$, the loss function $\mathcal{L}$, the number of samples $n$, the training set $\{(\ket{\mathbf{x}_m}, \mathbf{a}_m)\}_{m=1}^n$, the batch size $n_b$, the number of iterations $T$, the learning rate $\epsilon$, and the Adam optimizer $f_{\text{Adam}}$
\ENSURE The trained model
\STATE Initialization: generate random initial parameters for $\boldsymbol{\theta}$
\FOR{ $i \in [T]$}
    \STATE Divide the $150$ variational parameters into $10$ parameter-batches $\{b_1,b_2,...,b_{10}\}$, with each parameter-batch denoting the parameters encoded on the same qubit (i.e., the same row in the QNN circuit)
    \FOR{ $j \in [10]$}
        \STATE Randomly choose $n_b$ samples $\{\ket{\mathbf{x}_{\text{(i,j,1)}}},\ket{\mathbf{x}_{\text{(i,j,2)}}},...,\ket{\mathbf{x}_{\mathrm{(i,j,n_b)}}}\}$ among the $n$ samples in the training set
        \STATE Calculate the gradients for parameter-batch $b_j$ in experiments using the ``parameter shift rule'', and take the average value over the training batch $\mathbf{G}\leftarrow\frac{1}{n_b}\Sigma_{k=1}^{n_b}\nabla \mathcal{L}(h(\ket{\mathbf{x}_{\text{(i,j,k)}}};b_j),\mathbf{a}_{\text{(i,j,k)}})$
        \STATE Updates: $b_j \leftarrow f_{\text{Adam}}(b_j,\epsilon,\mathbf{G})$
    \ENDFOR
\ENDFOR
\STATE Output the trained model
\end{algorithmic}
\end{algorithm}
\end{figure}

\begin{figure}
\begin{algorithm}[H]
\caption{Generating adversarial examples for the quantum data}
\label{Algo_Adv_Qdata}
\begin{algorithmic}[1]
\REQUIRE The model $h$ with trained parameters $\boldsymbol{\theta}^{*}$, the loss function $\mathcal{L}$, the number of iterations $T$, the learning rate $\epsilon$, the Adam optimizer $f_{\text{Adam}}$, a coefficient $\kappa$ to control the range of the perturbation angles in the single qubit gates, and a legitimate sample $\ket{\mathbf{x}}$ with label $\mathbf{a}$
\REQUIRE Prepare a perturbation layer $U_{\boldsymbol{\psi}}$ which only contains single-qubit perturbations and all elements in $\boldsymbol{\psi}$ are initialized to zero such that the initial perturbation layer is equal to an identity operator, and an element $\boldsymbol{\psi}_i$ is mapped to a single-qubit rotation angle by $\kappa \sin{\boldsymbol{\psi}_i}$ such that the range of the perturbation angles in the single qubit gates can be upper bounded by $\kappa$
\ENSURE The adversarial example $\ket{\mathbf{x}^{\mathrm{adv}}}$
\STATE Initialization: Prepare the N\'{e}el state $\ket{\mathrm{N}}$ and denote the evolution under the Hamiltonian of the AA model as $U_{\mathrm{AA}}$
\FOR{ $i \in [T]$}
    \STATE Calculate the gradients of the loss function $\mathcal{L}$ with respect to the parameters in the perturbation layer \\ $\mathbf{G}_{\boldsymbol{\psi}}\leftarrow\nabla_{\boldsymbol{\psi}} \mathcal{L}(h(U_{\mathrm{AA}}U_{\boldsymbol{\psi}}\ket{\mathrm{N}};\boldsymbol{\theta}^{*}),\mathbf{a})$
    \STATE Updates: $\boldsymbol{\psi} \leftarrow f_{\text{Adam}}(\boldsymbol{\psi},\epsilon,-\mathbf{G}_{\boldsymbol{\psi}})$
\ENDFOR
\STATE Output $\ket{\mathbf{x}^{\mathrm{adv}}} \leftarrow U_{\mathrm{AA}}U_{\boldsymbol{\psi}}\ket{\mathrm{N}}$
\end{algorithmic}
\end{algorithm}
\end{figure}

% \begin{figure*}[hltp]
% \center
% \includegraphics[width=0.6\linewidth]{IPR.pdf}
% \caption{\textbf{The inverse participation ratio (IPR) with respect to $V/g$,
% where a phase transition occurs at the point $V/g = 2$.}
% For each $V/g$ point, we sample $1000$ times and take the average IPR value for illustration.}
% \label{fig:ipr}
% \end{figure*}

\subsection{Adversarial examples}
\label{app:adv_example}

As illustrated in the main text, the trained QNN classifier is able to classify the localized and thermal states with decent accuracy.
Furthermore, our goal is to generate adversarial examples that keep the original states' property while lead the classifier to make incorrect predictions.
To achieve this, we choose to design local perturbations during the state preparation process.
For concreteness,
the legitimate quantum data is generated by preparing the system to the N\'{e}el state and steering the system to evolve under the Hamiltonian of the AA model.
For the adversarial data, after preparing the system to the N\'{e}el state, we add local perturbations to each qubit and then continue the steering process. 
These perturbations are initially set as identity operators and contain parameters that can be optimized to maximize the loss function.
The strategy for generating adversarial examples is summarized in Algorithm \ref{Algo_Adv_Qdata} with the experimental performance exhibited in the main text.
% In addition, to demonstrate that there is a phase transition at the point $V/g = 2$,
% i.e., the two classes of state we choose in the main text are distinct to each other,
% we numerically calculate the inverse participation ratio (IPR) as shown in Fig.~\ref{fig:ipr} to make our work more convincing.

\section{Experimental details \label{sec:exp}}
\label{app:exp}

\subsection{Device information}

\begin{figure*}[h]
\center
\includegraphics[width=0.8\linewidth]{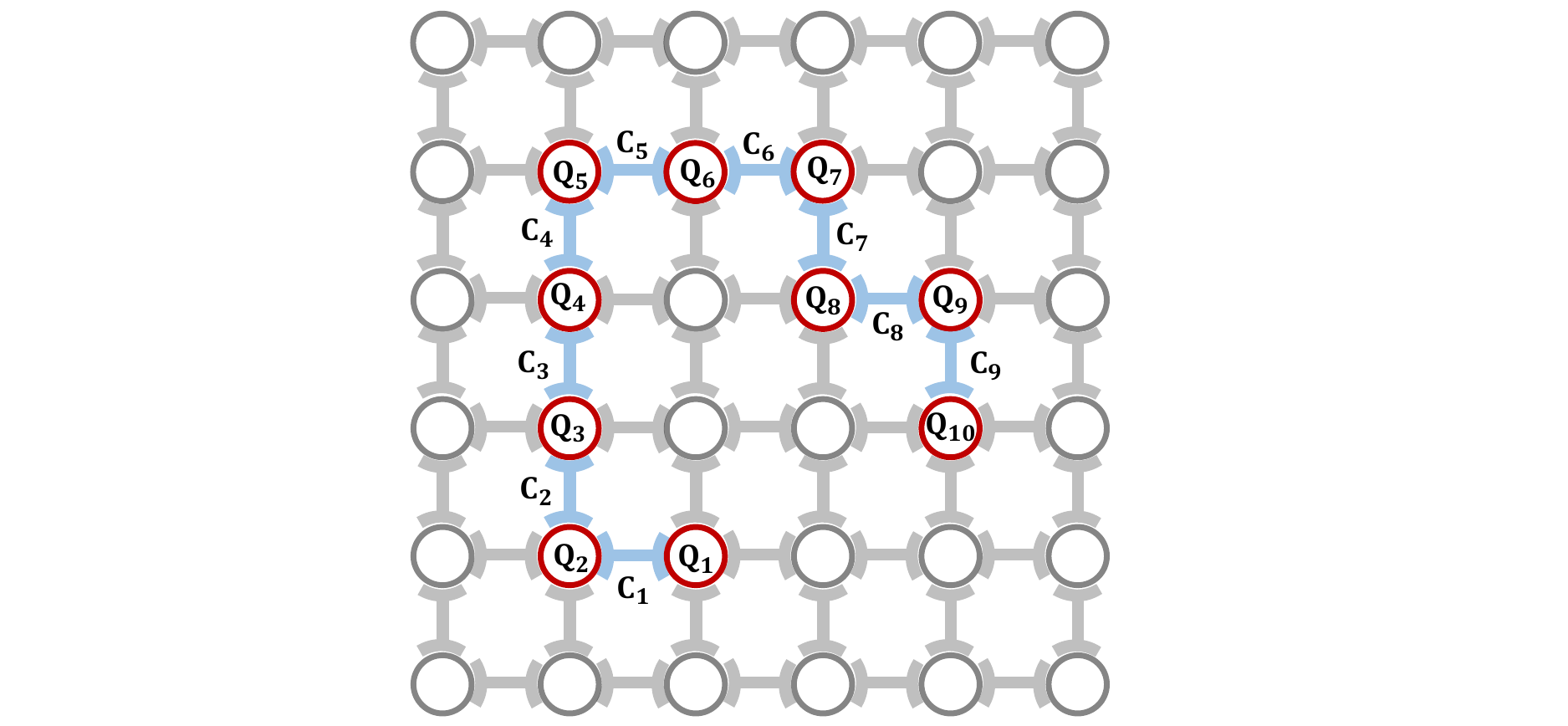}
\caption{\textbf{Layout of the multi-qubit superconducting processor.} The $L$ ($= 10$) qubits and $L-1$ couplers used for the experiment are colored in red and blue, respectively.}
\label{fig:device}
\end{figure*}

Our experiment is performed on a multi-qubit superconducting processor, with $6 \times 6$ transmon qubits arranged in a square lattice and $60$ couplers each inserted inbetween neighboring two qubits. Each qubit has nonlinearity around $-210$~MHz, with individual microwave line for XY gates and flux line for frequency tunability and Z gates; each coupler is also a transmon qubit whose nonlinearity is around $-250$~MHz, with individual flux line for frequency adjustment in the range from $\sim$4 to 6.5 GHz that is critical for turning on and off the effective coupling between the neighboring two qubits. We use tantalum film to pattern base wirings for high coherence and details on the device fabrication can be found in Ref.~\cite{Wang2021Quantum_SPT_TC}. To realize the ``interleaved'' block-encoding QNN structure, we select a chain of $L$ ($= 10$) qubits, Q$_j$ where $j = 1, 2, \ldots, 10$, as shown in Fig.~\ref{fig:device}. Characteristic parameters for these $L$ qubits are listed in Tab.~\ref{tab:deviceParameters}.

\begin{table*}[tbp]
\centering
\caption{\textbf{Characteristic device parameters.} %$\omega_{j}^r$ is the readout frequency of Q$_{j}$.
$\omega_{j}^0$ is the idle frequency where Q$_j$ is initialized and operated with the single-qubit rotational gates.
Nonlinearity $\eta_j$ of Q$_j$ is defined as the frequency difference between the $|1\rangle $-$|2\rangle$ and $|0\rangle $-$|1\rangle$ transitions.
$T_{1, j}$ is the energy relaxation time measured for Q$_j$ around $\omega_{j}^0$ and $T_{2,j}^{\text{DD}}$ is the dynamical decoupling (DD) dephasing time~\cite{PhysRevLett.121.130501} of Q$_j$ at $\omega_{j}^0$.
%$T_{1, j}$ and $T_{2,j}^{\text{DD}}$ are the energy relaxation time and dynamical decoupling (DD) dephasing time~\cite{PhysRevLett.121.130501} of Q$_j$ at $\omega_{j}^0$, respectively.
$F_{0, j}$ and $F_{1, j}$ are the readout fidelity values for Q$_j$ prepared in $|0\rangle$ and $|1\rangle$, respectively; these fidelity values are used to correct raw probabilities to eliminate
readout errors as done previously~\cite{PhysRevLett.118.210504}.
$\omega_{ij}^{\text{A (B)}}$ lists the estimated frequency values for Q$_i$ and Q$_j$ in group A (B) at which the CZ gate is implemented.
Pauli errors of the single-qubit gates ($e_1$) and those of the two-qubit CZ gates ($e_2^\text{A (B)}$) are characterized via simultaneous cross entropy benchmarking.}
\begin{center}
\begin{tabular}{p{5.0cm}<{\centering}p{0.9cm}<{\centering}p{0.9cm}<{\centering}p{0.9cm}<{\centering}p{0.9cm}<{\centering}p{0.9cm}<{\centering}p{0.9cm}<{\centering}p{0.9cm}<{\centering}p{0.9cm}<{\centering}p{0.9cm}<{\centering}p{0.9cm}<{\centering}p{0.01cm}<{\centering}|p{0.9cm}<{\centering}}
        \hline
        \hline
        %Qubit & $Q_{3_5}$ & $Q_{3_3}$ & $Q_{5_3}$ & %$Q_{7_3}$ & $Q_{9_3}$ & $Q_{9_5}$ & $Q_{9_7}$ %& $Q_{7_7}$ & $Q_{7_9}$ & $Q_{5_9}$ & Mean\\
        Qubit & Q$_{1}$ & Q$_{2}$ & Q$_{3}$ & Q$_{4}$ & Q$_{5}$ & Q$_{6}$ & Q$_{7}$ & Q$_{8}$ & Q$_{9}$ & Q$_{10}$ & & Mean\\
        \hline
        \hline
        $\omega_{j}^0/2\pi$ (\text{GHz}) & 4.260 & 4.390 & 4.545 & 4.690 & 4.380 & 4.280 & 4.120 & 4.400 & 4.250 & 3.990 & \\
        \hline
        $\eta_{j}/2\pi$ (\text{GHz}) & -216 & -213 & -211 & -209 & -213 & -213 & -216 & -212 & -216 & -220 & & -214\\
        \hline
        $T_{1, j}$ (\text{$\mu$s}) & 153 & 141 & 131 & 132 & 159 & 172 & 173 & 158 & 153 & 152 & & 152\\
        %$T_{1, j}$ (\text{$\mu$s}) & 164 & 133 & 115 & 111 & 150 & 164 & 178 & 145 & 158 & 139 & &146\\
        \hline
        $T_{2,j}^{\text{DD}}$ (\text{$\mu$s})& 91 & 54 & 121 & 105 & 95 & 93 & 99 & 127 & 143 & 75 & &100\\
        \hline
        $F_{0, j}$  & 0.976 & 0.976 & 0.993 & 0.994 & 0.987 & 0.990 & 0.977 & 0.989 & 0.990 & 0.983 & & 0.986\\
        \hline
        $F_{1, j}$  & 0.944 & 0.960 & 0.983 & 0.984 & 0.979 & 0.972 & 0.947 & 0.972 & 0.963 & 0.967 & & 0.967\\
        \hline
        \hline
        simultaneous 1Q XEB $e_1$ ($\%$) &  0.06 & 0.07 & 0.09 & 0.07 & 0.06 & 0.11 & 0.09 & 0.08 & 0.08 & 0.06 & & 0.08 \\
        \hline
        \hline
        $\omega_{ij}^{\text{A}}/2\pi$ (\text{GHz}) & \mcol{\cfill4.260,    4.465} &   \mcol{\cfill4.580,    4.781} & \mcol{\cfill4.430,    4.225} & \mcol{\cfill4.175,    4.378} & \mcol{\cfill4.235,    4.030}  & &\\
        \cline{1-12}
        $\omega_{ij}^{\text{B}}/2\pi$ (\text{GHz}) & & \mcol{\cfillo 4.370,    4.574} &   \mcol{\cfillo 4.591,    4.390} & \mcol{\cfillo 4.271,    4.065} & \mcol{\cfillo 4.403,    4.200} && \\  
        \hline
        simultaneous 2Q XEB $e_{2}^{\text{A}}$ ($\%$) & \mcol{\cfill 0.52} & \mcol{\cfill 0.65} & \mcol{\cfill 0.72} & \mcol{\cfill 0.77} & \mcol{\cfill 0.71} & &\multirow{2}{*}{0.72} \\
        \cline{1-12}
        simultaneous 2Q XEB $e_{2}^{\text{B}}$ ($\%$) & & \mcol{\cfillo 0.88} & \mcol{\cfillo 0.74} & \mcol{\cfillo 0.86} & \mcol{\cfillo 0.64} & &  \\
        \hline
        \hline
\end{tabular}
\end{center}
\label{tab:deviceParameters}
\end{table*}

\subsection{Experiment circuit}

\begin{figure*}[tbp]
	\center
	\includegraphics[width=1.0\linewidth]{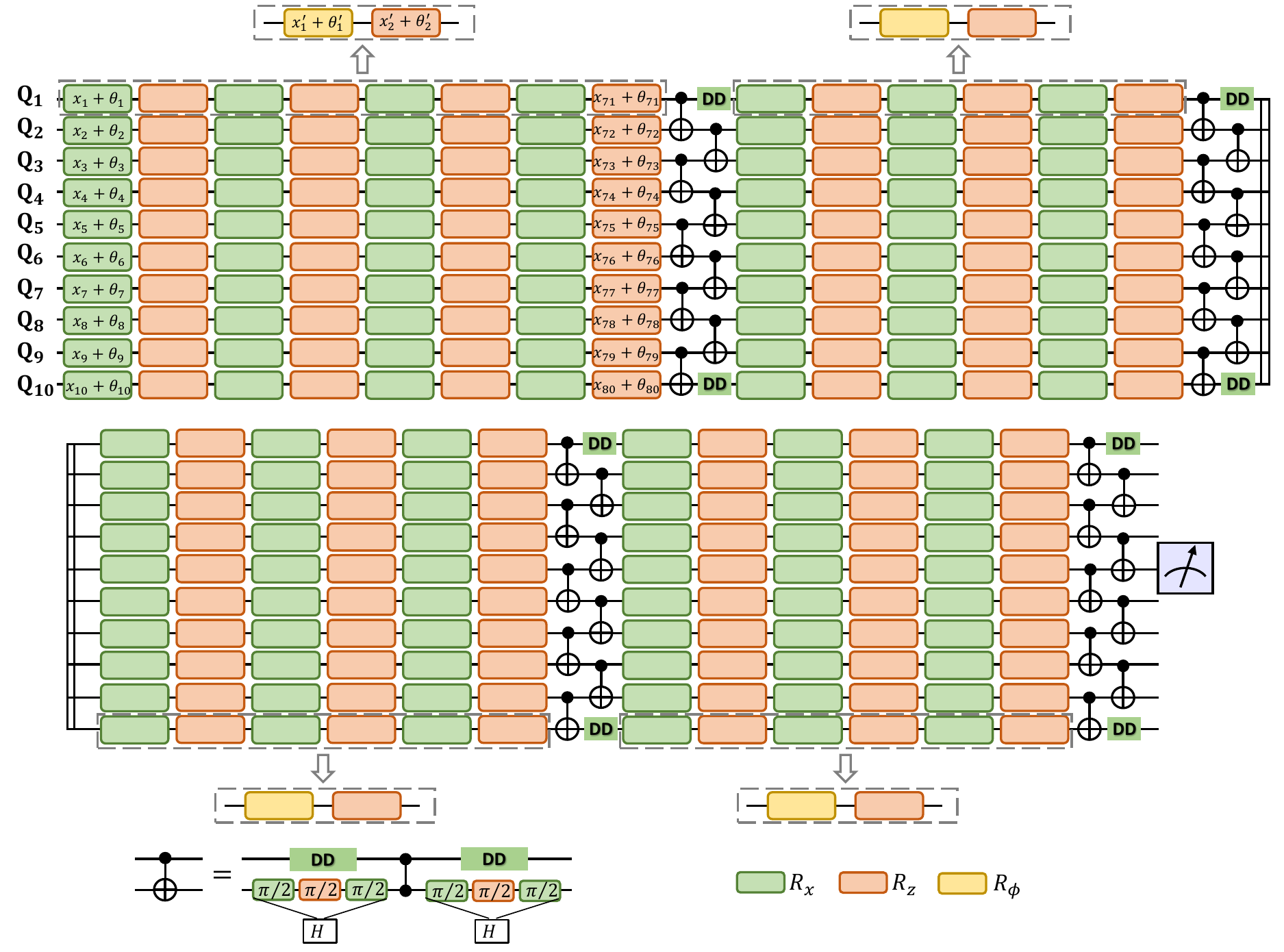}
	\caption{\textbf{Experiment circuit to encode classical data and realize the quantum neural network classifier for leaning medical data.} We apply dynamical decoupling (DD) sequences to the qubits that are idling to elongate the effective dephasing times.}
	\label{fig:circuit}
\end{figure*}

A fundamental QNN block for the chain topology is shown in Fig.~\ref{fig:circuit}, which consists of multiple layers of simultaneous single-qubit rotational gates, followed by two layers of CNOT gates running through $L-1$ neighboring qubit pairs. The single-qubit rotational (XY and Z) gates include $R_x$, $R_y$, and $R_z$, which rotate the qubit state by arbitrary angles around $x$-, $y$- and $z$-axis, respectively. The CNOT gate is composed of a generic two-qubit controlled $\pi$-phase (CZ) gate sandwiched inbetween two Hadamard gates, the later of which are realized by three single-qubit rotations $R_x(\pi/2) R_z(\pi/2) R_x(\pi/2)$. Experimentally, we initialize the $L$ qubits to the ground state at their respective idle frequencies $\omega_{j}^0$, where all single-qubit rotational gates are applied. When necessary, we bias qubit pairs to the frequency values listed in either $\omega_{ij}^{\text{A}}$ or $\omega_{ij}^{\text{B}}$ for CZ gates, where the superscripts A/B refers to the group of qubit pairs whose CZ gates are implemented in parallel. While running multiple CZ gates in parallel, we apply dynamical decoupling  sequences (see green boxes labeled as ``DD'' in Fig.~\ref{fig:circuit}) featuring two segments of microwave drives with opposite phases to the qubits that are idling, elongating the effective dephasing times of these qubits~\cite{PhysRevLett.121.130501}.

To encode the classical medical data which are pictures of $16\times 16$ grayscale pixels, we repeat the fundamental block 4 times to construct a variant of the QNN classifier (Fig.~\ref{fig:circuit}). The first block (each of the rest 3 blocks) contains $10\times 8$ ($10 \times 6$) single-qubit rotational gates selected from \{$R_x$, $R_z$\}, so that the QNN classifier can encode up to 260 rotation angle parameters which sufficiently cover the components of a normalized vector \textbf{x} converted from $16\times 16$ grayscale pixels, with the unused angle parameters preset to zero. In addition, in this variant the data-encoding blocks and variational blocks are merged together, i.e., the input \textbf{x} and trainable parameters $\boldsymbol{\theta}$ are summed up with certain weights as the input parameters for the rotation angles.

To further reduce the circuit depth for the experiment, as illustrated by dashed line boxes in Fig.~\ref{fig:circuit}, consecutive single-qubit gates are compiled and replaced by two single-qubit rotations $R_z\left(\theta\right) R_{\phi}\left(\theta^\prime\right)$ featuring three independent parameters $\theta$, $\theta^\prime$ and $\phi$, where the subscript $\phi$ refers to an equatorial rotation axis that has an angle $\phi$ with respect to $x$-axis.

\begin{figure*}[tbp]
	\center
	\includegraphics[width=1.0\linewidth]{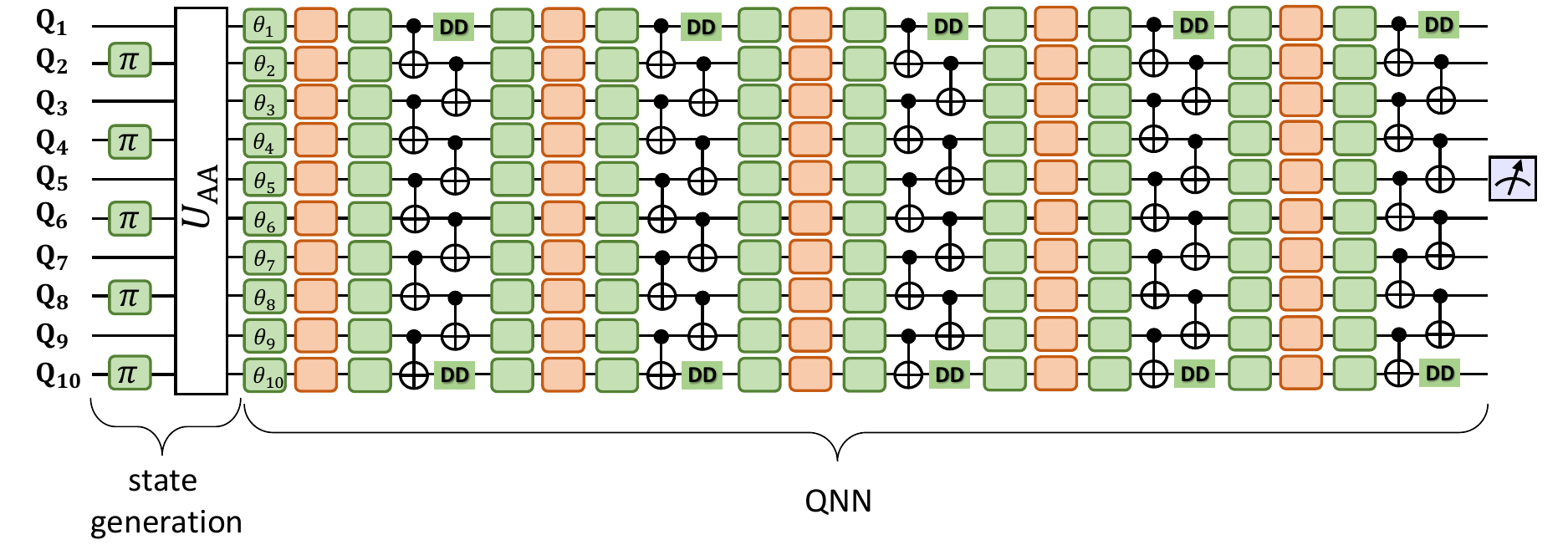}
	\caption{\textbf{Experiment circuit to generate quantum states and realize the quantum neural network classifier for quantum data training.} Here $U_{\text{AA}}=e^{-iH\tau}$ represents the evolution under the Aubry-Andr\'{e} (AA) Hamiltonian (Eq. 3 in the main text) for a fixed time $\tau=400$ ns.
	}
	\label{fig:circuit_q}
\end{figure*}

Since the input state already encodes the data, the QNN classifier for quantum data training employs 5 variational blocks with $10 \times 3$ single-qubit gates (30 training parameters) in each block, which amount to 150 training parameters as shown in Fig.~\ref{fig:circuit_q}. Below we focus on characterizing the single- and two-qubit gates, which are the most critical elements required in the QNN classifiers.

\subsubsection{Single-qubit gates}
The single-qubit XY gates ($R_x$ and $R_y$) are realized by 30 ns-long microwave pulses with Gaussian envelopes, where the quadrature correction terms with DRAG coefficients are implemented to minimize state leakage to higher levels~\cite{PhysRevLett.112.240504}. 
Due to the existence of microwave crosstalk, during the implementation of random XY gates on multiple qubits simultaneously, 
individual qubits are susceptible to the off-resonant microwave pulses applied to the drive lines that are designed to address other qubits, 
necessitating an active microwave cancellation technique~\cite{xukai2020, PhysRevX.11.021058, Wallraff_qec}. 
Here we quantify the microwave crosstalk with a complex matrix $M$ defined as $\tilde{\Omega}_\text{actual} = M\cdot\tilde{\Omega}_\text{applied}$, where $\tilde{\Omega}_\text{applied}$ ($\tilde{\Omega}_\text{actual}$) is a column vector containing the microwave tones applied to (actually sensed by) all the qubits. Suppose we apply on Q${_j}$ a microwave tone $\Omega_{j}(t)$, the tone sensed by Q$_{k}$ due to the crosstalk effect can be written as $\Omega_{k}(t) = M_{kj}\Omega_{j}(t)$, where $M_{kj}=A_{kj}e^{i\varphi_{kj}}$ is a complex factor with $A_{kj}$ and $\varphi_{kj}$ being the crosstalk amplitude and phase, respectively.
To characterize $A_{kj}$ and $\varphi_{kj}$, we use the sequence fidelity of Q$_k$ in randomized benchmarking (RB) as a fitness metric while Q$_j$ is also subject to RB pulses,
and find that the microwave crosstalk matrix $M$ is sparse and has an average amplitude of around 4\% among the non-zero off-diagonal terms.

Based on the calibrated microwave crosstalk matrix $M$ and the active microwave cancellation technique, we have verified via cross entropy benchmarking (see below) 
that our single-qubit XY Pauli errors, $e_1$, are averaged to be 0.08\% 
for the case of implementing random XY gates on all 10 qubits simultaneously (Tab.~\ref{tab:deviceParameters}).

The single-qubit Z gates ($R_z$) are mostly implemented via virtual Z gates~\cite{PhysRevA.96.022330} in the QNN classifiers.
We have also benchmarked $R_z(\pi/2)$ realized by 20 ns-long square pulses, yielding an average Pauli error of 0.03\% for the case of simultaneously running on all 10 qubits,
which is better than that for the XY gates as microwave crosstalk is not a concern here.

\subsubsection{Two-qubit CZ gate}
\begin{figure*}[h]
	\center
	\includegraphics[width=1\linewidth]{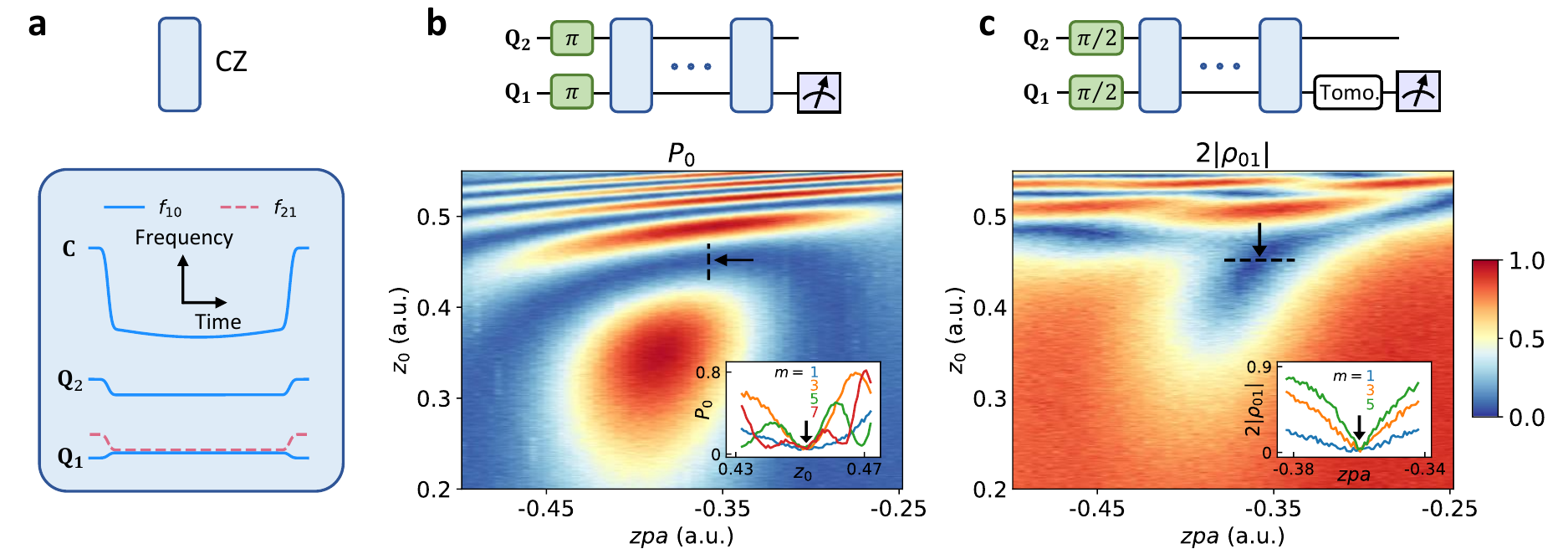}
	\caption{\textbf{Tuning up the CZ gate for Q$_1$ and Q$_2$.} \textbf{a}, CZ pulse sequence plotted in the frequency versus time domain. \textbf{b}, $|0\rangle$-state probability of Q$_{1}$, $P_0$, as function of $zpa$ and $z_0$ at $m=1$. Inset shows one-dimensional sweeps of $P_0$ vs. $z_0$ at different $m$ values along the dashed line in the main panel.
		\textbf{c}, Off-diagonal $|\rho_{01}|$ of Q$_{1}$'s density matrix as function of $zpa$ and $z_0$ at $m=1$. Inset shows one-dimensional sweeps of $|\rho_{01}|$ vs. $zpa$ at different $m$ values along the dashed line in the main panel. Corresponding experimental sequences are shown on top for data in \textbf{b} and \textbf{c}.
	}
	\label{fig:tune_cz}
\end{figure*}

The CZ gate between two neighboring qubits Q$_j$ and Q$_{j+1}$ is realized by dynamically steering the resonant frequency of the coupler, C,
along a well-designed trajectory, so that the effective coupling strength can be turned on for a specific amount of time. Here we describe
the procedure of parametrizing the CZ process to maximize its gate fidelity. We use the notation $|\text{Q}_j, \text{C}, \text{Q}_{j+1}\rangle$
to represent the dressed state of the three-body system, and assume
that Q$_j$ has the lowest frequency. During the CZ process, small square flux (z) pulses are applied to Q$_j$ and Q$_{j+1}$
so that $|101\rangle$ and $|002\rangle$ are near resonance, while a sine decorated square flux (z) pulse with the form
\begin{equation}
	z(t) = z_0\left[1-r+r\sin{\left(\pi\frac{t}{t_{\text{gate}}}\right)}\right]
	\label{ripple}
\end{equation}
is applied to C to lower its frequency from $\sim$5.8~GHz to $\sim$4.6~GHz, where $t_{\text{gate}}=50$~ns and $r$ is an optimization parameter typically $\sim$0.1.
We add 5 ns zero-paddings before and after $t_{\text{gate}}$ when concatenating gates. Here $z_0$ is tuned to minimize the leakage from $|101\rangle$ to $|002\rangle$, while the z pulse amplitude ($zpa$) of Q$_{j+1}$ (or Q$_{j}$) is adjusted to maximize qubit entanglement.
Below we illustrate a couple of key steps during the repeated optimization process (see Fig.~\ref{fig:tune_cz}):
\begin{enumerate}
	\item Optimizing $z_0$: After coarse adjustment of all parameters,
	we prepare $|101\rangle$, run the CZ pulses for $m$ cycles with $m \in \{1, 3, 5, 7\}$, and finally measure the $|0\rangle$-state probability of Q$_{j}$, $P_0$.
	We identify the optimal $z_0$ at which the averaged $P_0$ reaches minimum, indicating the lowest state leakage (see Fig.~\ref{fig:tune_cz}\textbf{b} and inset).
	\item Optimizing $zpa$ of Q$_{j+1}$ (or Q$_j$): With $z_0$ obtained from step 1,
	we prepare both qubits in $\left(|0\rangle - i|1\rangle\right)/\sqrt{2}$ and the coupler in $|0\rangle$, run the CZ pulses for $m$ cycles with $m \in \{1, 3, 5\}$, and finally perform tomographic measurement
	on Q$_{j}$ to extract the off-diagonal $\rho_{01}$ of its density matrix. We identify the optimal $zpa$ of Q$_{j+1}$ (or Q$_j$) at which the averaged $|\rho_{01}|$ reaches minimum (see Fig.~\ref{fig:tune_cz}\textbf{c} and inset).
\end{enumerate}

\subsubsection{Quantum gate benchmarks}

\begin{figure*}[tbp]
	\center
	\includegraphics[width=0.8\linewidth]{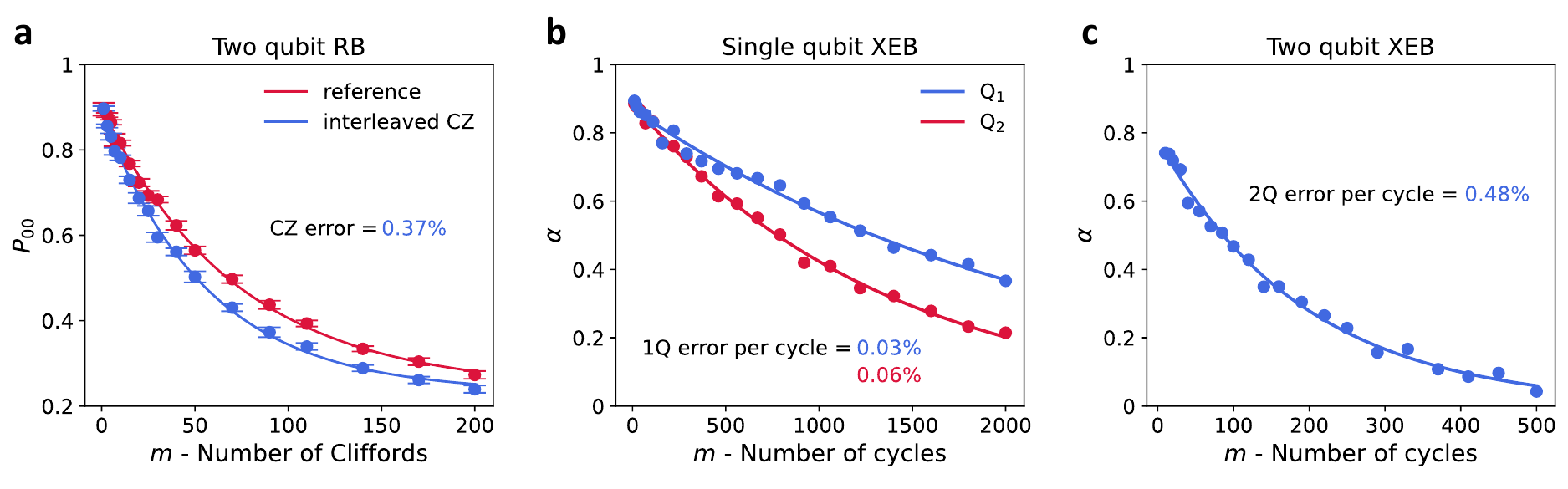}
	\caption{\textbf{Comparison of XEB and RB in benchmarking a two-qubit CZ gate.}
		\textbf{a}, Interleaved RB data to characterize the CZ gate on Q$_1$ and Q$_2$.
		\textbf{b}, XEB data taken simultaneously on Q$_1$ and Q$_2$ to characterize the single-qubit gates.
		\textbf{c}, XEB data to characterize the CZ gate on Q$_1$ and Q$_2$, where each cycle contains two single-qubit gates in parallel and a CZ gate.
		The CZ Pauli errors extracted from RB and XEB are $0.37\%$ and $0.39\%$, respectively, which are consistent.}
	\label{fig:fig_sanity}
\end{figure*}

Here we focus on characterizing performance of the quantum gates via cross entropy benchmarking (XEB)~\cite{boixo_characterizing_2018,Arute2019Quantum}, in particular when these gates are implemented on multiple qubits simultaneously. We verify that both XEB and RB yield very similar error values in benchmarking our experimental gates, with an example shown in Fig.~\ref{fig:fig_sanity}.

\begin{figure*}[tbp]
	\includegraphics[width=1.0\linewidth]{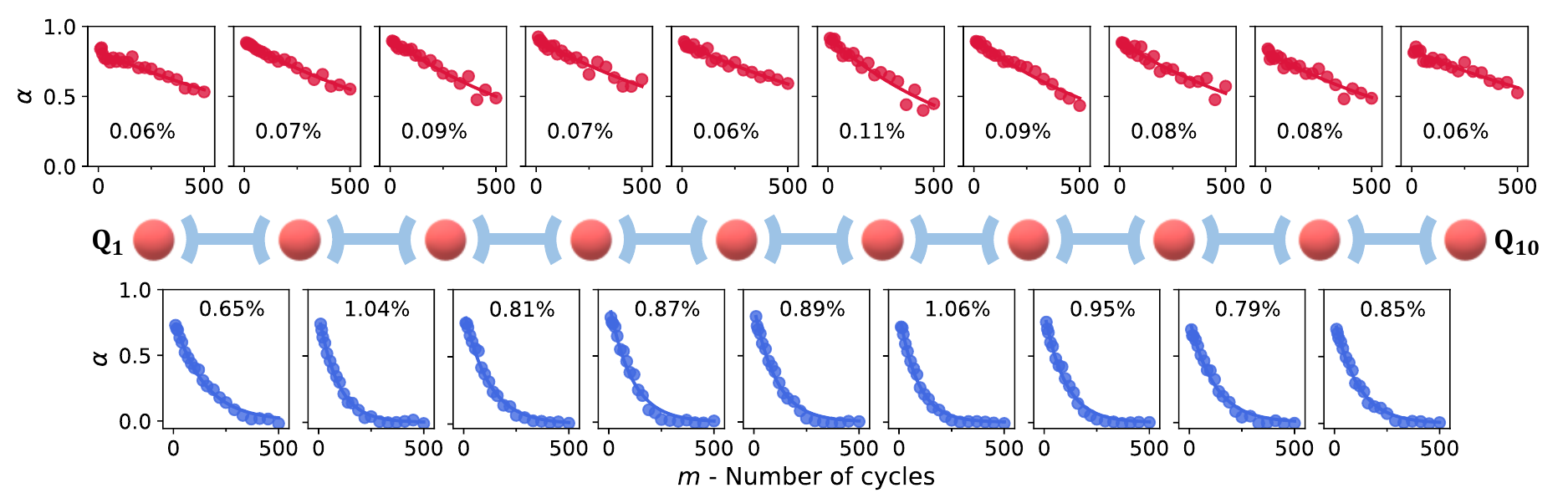}
	\center
	\caption{\textbf{Simultaneous XEB results of the single-qubit gates and the two-qubit CZ gates.}
		We sample over 50 different random circuits to calculate the sequence fidelity $\alpha$ as a function of cycle number $m$ (circles) which is fitted to $\alpha = Ap^{m}$ (lines). The Pauli error per cycle $e_{c}$ illustrated in each figure is calculated by $\left(1-p\right)\left(1-1/D^2\right)$.
		For the single-qubit gates, the Pauli errors per gate $e_1$ in Tab.~\ref{tab:deviceParameters} are just $e_{c}$ as listed in the figures.
		For the two-qubit CZ gates, the Pauli errors per gate $e_2$ in Tab.~\ref{tab:deviceParameters} are calculated according to $\left(1-e_c\right) = \left(1-e_{c, j}\right)\left(1-e_{c, {j+1}}\right)\left(1-e_2\right)$, where $e_{c, j}$ ($e_{c, j+1}$) refers to the Pauli error for Q$_j$ (Q$_{j+1}$).}
\label{fig:fig_xeb}
\end{figure*}

Simultaneous XEB results characterizing the quantum gates are shown in Fig.~\ref{fig:fig_xeb}. For single-qubit gates, each cycle in an XEB circuit consists of a $\pi/2$ rotation randomly chosen from the following set: $R_{\phi}\left(\frac{\pi}{2}\right)$ where $\phi \in  \{0, \frac{1}{4}\pi, \frac{1}{2}\pi, \frac{3}{4}\pi, \pi, \frac{5}{4}\pi, \frac{3}{2}\pi, \frac{7}{4}\pi\}$.
At the end of the circuit, a random single-qubit gate $R_{\phi}\left(\theta\right)$ is applied to randomize the circuit and achieve Porter-Thomas distribution required for XEB~\cite{Arute2019Quantum} with $\phi$ and $\theta$ subjected to the probability density function
\begin{equation}
f(\phi, \theta)=\frac{1}{4\pi}\sin{\theta}. %\theta\in[0, \pi], \phi\in[0, 2\pi).
\end{equation}

For two-qubit CZ gates, the single-qubit gate set used in each cycle is the same as the one mentioned above, and each cycle contains a layer of two single-qubit gates followed by a CZ gate. Similarly, to approach Porter-Thomas distribution more quickly, every XEB circuit ends with a layer of random single-qubit gates.
We can calculate the sequence fidelity $\alpha$ with the measured probabilities of bitstrings using the following relation
\begin{equation}
\sum_{q\in\{0,1\}^n} \overline{p_{e}\left(q\right)\left(D p_{s}\left(q\right)-1\right)} = \alpha\left(D \sum_{q\in\{0,1\}^n} \overline{p_{s}\left(q\right)^{2}} - 1\right),
\label{XEB}
\end{equation}
where D ($= 2^n$) is the dimension of Hilbert space, $p_{s}(q)$ and $p_{e}(q)$ are the simulated and experimentally measured probabilities of bitstring $q$, respectively, and the horizontal bar on the top denotes averaging over random circuits. We note that XEB can be used to further tune up the CZ gates.

\subsection{MNIST data training}

\begin{figure}[tbp]
	\center
	\includegraphics[width=1.0\linewidth]{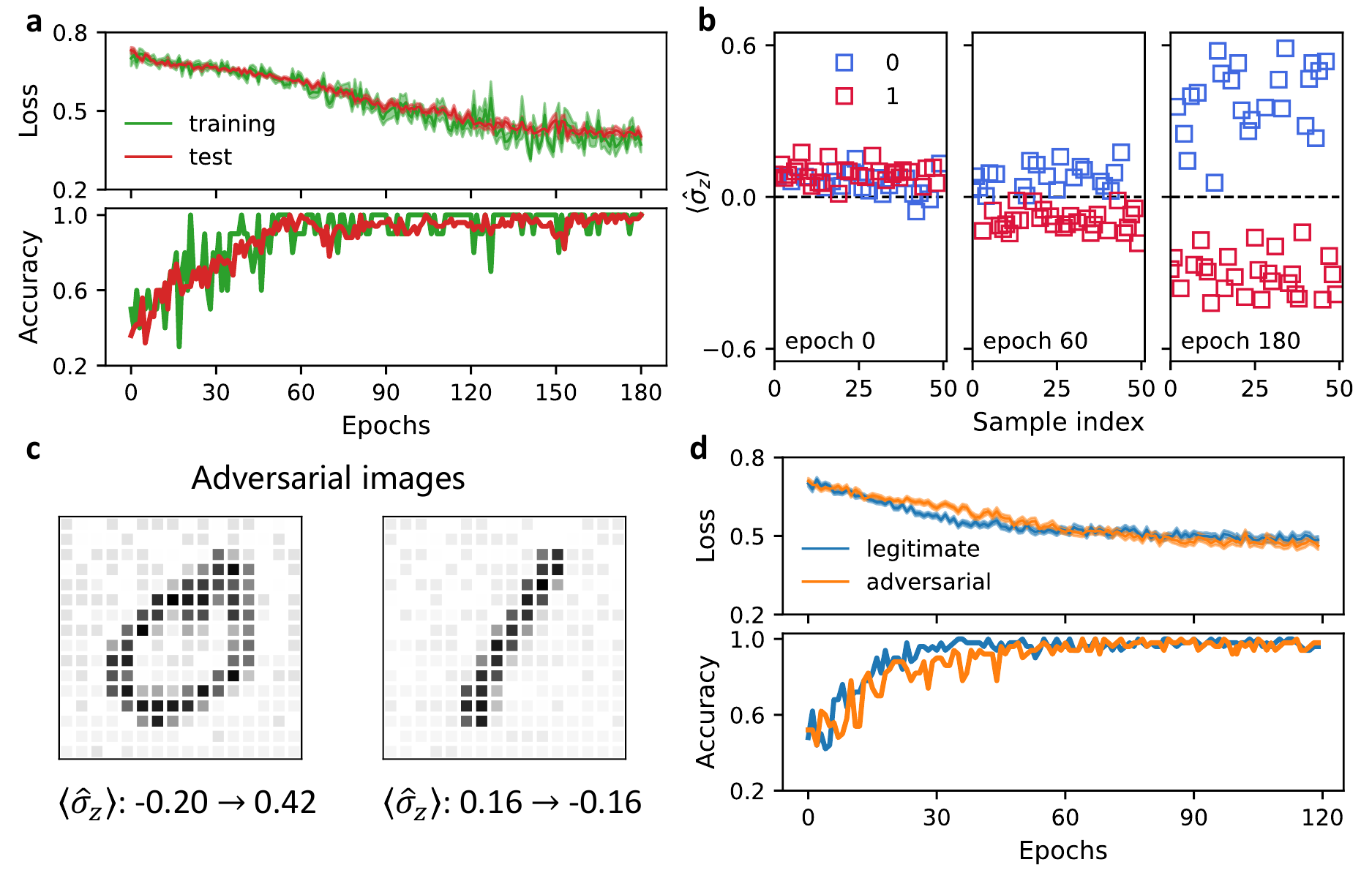}
	\caption{\textbf{Experimental results of the MNIST data training and adversarial quantum machine learning.} \textbf{a}, Loss function (up) and accuracy (down) for the training and testing data set at each epoch. \textbf{b}, Experimentally measured $\langle\hat\sigma_z\rangle$ of Q$_5$ for the test data at epoch $0$, $60$ and $180$. Data for digits ``0'' and ``1'' are colored in blue and red, respectively. \textbf{c}, Legitimate and adversarial samples with measured output $\langle\hat{\sigma}_z\rangle$ for Q$_5$ of the trained quantum classifier. \textbf{d}, Loss function (up) and accuracy (down) for the legitimate and adversarial test data at each epoch.}
	\label{fig:MNIST}
\end{figure}

We select ``0" and ``1" from the MNIST digits to form the training and test data sets, with sample sizes of $500$ and $100$, respectively. We start the training by assigning the QNN classifier with randomly generated trainable parameters. At each epoch, we select $10$ ($50$) digits to form the training (test) data randomly. While there are $260$ trainable parameters (Fig.~\ref{fig:circuit}), we find that a reduced number of parameters is enough to train the classifier. In practice, we set the learning rate to $0.02$, and select $50$ parameters to train, while unused parameters are initialized to randomly assigned values and remain unchanged throughout the learning process. The experiment results are shown in Fig.~\ref{fig:MNIST}\textbf{a}. We plot the loss function and accuracy of both the training and test data measured at each epoch. As the loss function decreases slowly during the learning process, the accuracy increases at a relatively faster speed and approaches to $1$ after about $50$ epochs. Further decrease of the loss function helps to enhance the visibility of the classifier as witnessed by the instances in Fig.~\ref{fig:MNIST}\textbf{b}. The trained QNN classifier can classify the total training and test data sets accurately.

We also explore the behavior of the QNN classifier under the adversarial attacks. The adversarial samples are generated by adding a small but carefully-designed perturbation to the digits. In this work, the perturbation is designed by maximizing the loss function numerically, and the classifier indeed fails to classify resulted adversarial samples (see examples in Fig.~\ref{fig:MNIST}\textbf{c}). Next we include the adversarial samples and implement adversarial machine learning. We start the training by re-initialize the $50$ trainable parameters. At each epoch, we randomly select $5$ ($50$) samples from the original data set and $5$ ($50$) from the adversarial data set to form the training (test) data. The loss function and accuracy for both the legitimate and adversarial samples measured at each epoch are shown in Fig.~\ref{fig:MNIST}\textbf{d}. The re-trained classifier can defend certain adversarial attacks better than the original classifiers.

\end{document}